\documentclass{article}
\usepackage[top=20mm,left=20mm,right=20mm,bottom=20mm]{geometry}
\usepackage{authblk}
\usepackage{amsmath,amssymb}
\usepackage{subcaption}
\usepackage{graphicx}

\newcommand\ope{\mathrm{e}}
\newcommand\opd{\mathrm{d}}
\newcommand\opi{\mathrm{i}}
\newcommand\oppi{\mathrm{\pi}}
\def\vec#1{\mbox{\boldmath{\protect{\begin{math}#1\end{math}}}}}

\usepackage{multirow}
\usepackage{tikz}
\usetikzlibrary{arrows.meta}

\newcommand{\dif}{\mathop{}\!\mathrm{d}}
\newcommand{\Gd}{G_{\textrm{d}}}
\NewDocumentCommand\vecU{o}{\vec{U}\IfValueT{#1}{^{(#1)}}}
\NewDocumentCommand\vecV{o}{\vec{V}\IfValueT{#1}{^{(#1)}}}
\newcommand{\Uk}{U_k}
\newcommand{\Vk}{V_k}
\newcommand{\UU}[2]{U_{#1}^{(#2)}}
\newcommand{\VV}[2]{V_{#1}^{(#2)}}
\newcommand{\bigpw}[2]{\big({#1}\big)^{#2}}
\newcommand{\dkp}{\delta_{k}^{+}}
\newcommand{\dkm}{\delta_{k}^{-}}

\newcommand{\dktwo}{\delta_{k}^{\langle 2 \rangle}}
\newcommand{\largevec}[2]{
  \left(
    \begin{matrix}
      {#1} 
    \\
      {#2}
    \end{matrix}
  \right)
}
\newcommand{\largemat}[4]{
  \left(
    \begin{matrix}
      {#1} & {#2}
    \vspace{1mm}\\
      {#3} & {#4}
    \end{matrix}
  \right)
}

\newlength\rowht
\newlength\arrowpad
\newcommand{\darrow}[1]{%
  \multirow{#1}{*}{%
    \tikz[baseline]{%
      \draw[
        ->,
        >={Stealth[scale=1.2]},
        line width=0.6pt
      ]
      (0,\dimexpr 0.6\rowht \relax)
      --
      (0,\dimexpr 0.6\rowht - #1\rowht + 0.2\rowht\relax);
    }%
  }%
}

\makeindex

\begin{document}


\title{Exploring global landscape of free energy \\ for the coupled Cahn--Hilliard equations}

\author[1]{Keiichiro Kagawa}
\author[2]{Takeshi Watanabe}
\author[1,3,4]{Yasumasa Nishiura\footnote{yasumasa@pp.iij4u.or.jp}}
\affil[1]{Research Center of Mathematics for Social Creativity, Research Institute for Electronic Science, Hokkaido University, Kita 12 Nishi 7, Sapporo Hokkaido, 060-0812, Japan}
\affil[2]{Nagano University, 658-1 Shimonogo, Ueda Nagano, 386-1298, Japan}
\affil[3]{Advanced Institute for Materials Research (WPI-AIMR), Tohoku University, Sendai, Miyagi, 980-8577, Japan}
\affil[4]{Chubu University Academy of Emerging Sciences, Chubu University, Aichi, Kasugai, 487-8501, Japan}
\date{}

\maketitle

\begin{abstract}
Describing the complex landscape of infinite-dimensional free energy is generally a challenging problem. This difficulty arises from the existence of numerous minimizers and, consequently, a vast number of saddle points. These factors make it challenging to predict the location of desired configurations or to forecast the trajectories and pathways leading from an initial condition to the final state.
In contrast, experimental observations demonstrate that specific morphologies can be reproducibly obtained in high yield under controlled conditions, even amidst noise. This study investigates the possibility of elucidating the global structure of the free energy landscape and enabling the control of orbits toward desired minimizers without relying on exhaustive brute-force methods. Furthermore, it seeks to mathematically explain the efficacy of certain experimental setups in achieving high-yield outcomes.
Focusing on the phase separation of two polymers in a solvent, we conduct a one-dimensional analysis that reveals the global free energy landscape and relaxation-parameter-dependent trajectory behaviors. 
Two key methodologies are developed: one is a saddle point search method, akin to bifurcation tracking. This method aims to comprehensively identify all saddle points. The other is a strategy that adjusts the relaxation parameters preceding each variable's time derivative, aligning with experimental setups. This approach enables control over trajectory behaviors toward desired structures, overcoming the limitations of steepest descent methods. By tuning these relaxation parameters, uncertainties in trajectory behavior due to inevitable fluctuations can be suppressed. These methodologies collectively offer a mathematical framework that mirrors experimental high-yield phenomena, facilitating a deeper understanding of the underlying mechanisms.
\end{abstract}

\begin{center}
  {\bf Keywords} free energy, phase separation,
  structure preserving scheme,
  Cahn--Hilliard equation, bifurcated path tracking, block copolymer,
  nanoparticle
\end{center}


\section{Introduction}
\begin{figure}[h]
  \begin{center}
    \includegraphics[width=.5\hsize]{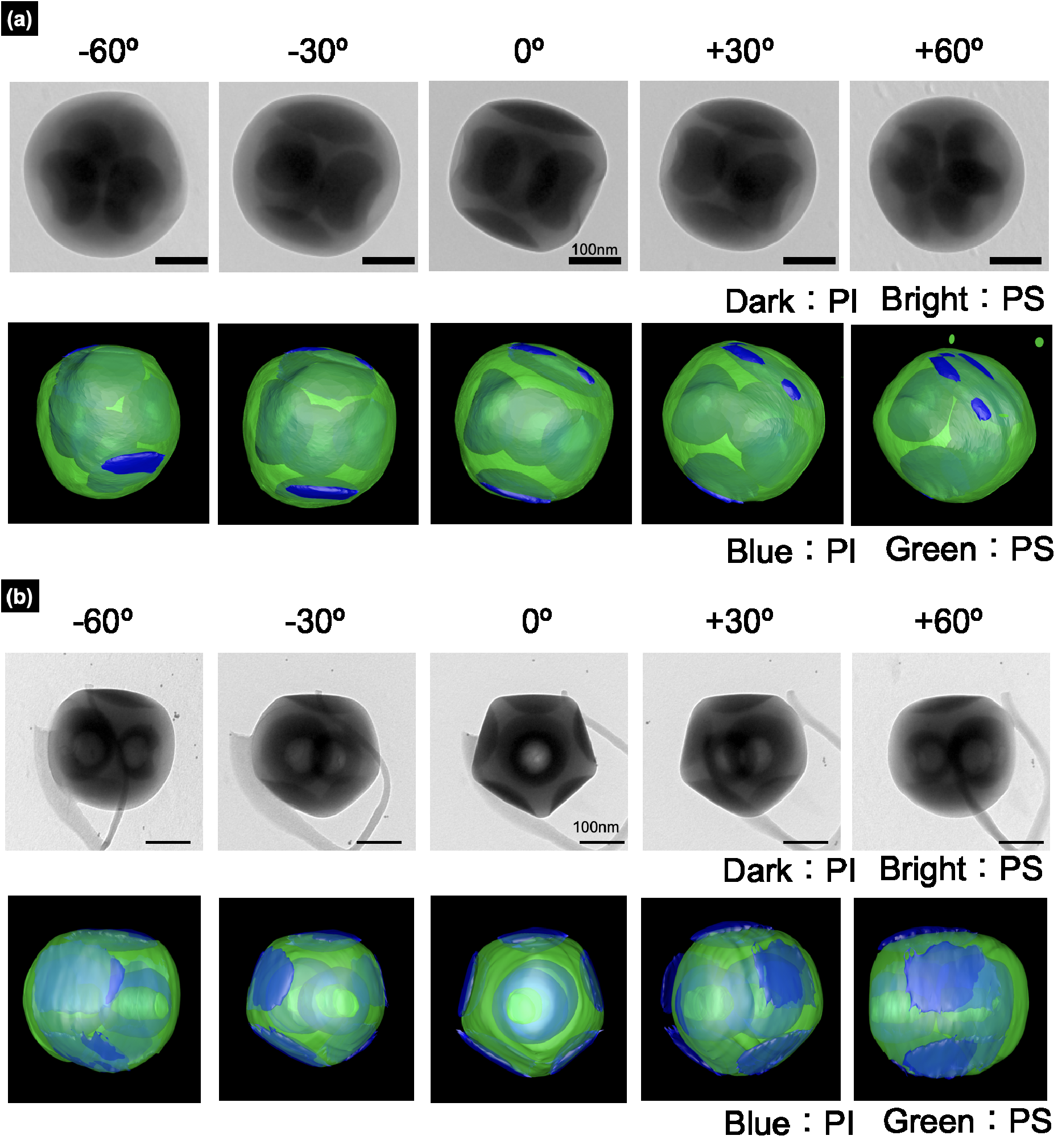}
  \end{center}
  \caption{
    Two polyhedral nanoparticles: (a) cubic-like and (b) octahedron-like. Tilted TEM images (upper row) and tilted ET images (bottom row) of PS-\textit{b}-PI particles with $D/L_0=3.8$ for (a), and $D/L_0=3.9$ for (b). Here $D/L_0$ is the ratio of diameter of the particle and the length of lamellar structure of PS-\textit{b}-PI in the bulk. The scale bar shows 100 nm. 
    In the TEM images, PI and PS phases are shown dark and bright contrast, and those are shown as blue and green phases in the ET images, respectively. Image courtesy of Dr. Hiroshi Yabu.
  }
  \label{fig:yabu}
\end{figure}

\begin{figure}[h]
  \begin{center}
    \includegraphics[width=.7\hsize]{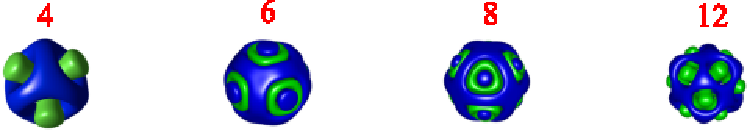}
  \end{center}
  \caption{
    Examples of some polyhedral particles obtained with the theoretical model (5). From left to right, the particles have 4, 6, 8 and  12 faces, respectively. Although the specific morphology depends on several parameters, roughly speaking the number of faces in this figure increases with the width of the interface of the $v$ component, The values of $\epsilon_v$ from left to right are, $0.02$, $0.02$, $0.0208$ and $0.036$. The relaxation parameters for the first and second particles are $\tau_v=10$ and $\tau_v=400$, whereas for the third and fourth cases $\tau_v=100$. In all cases $\tau_u=1$.
    The system size is $L=0.8$ for the first case and $L=1.0$ for the rest. Image courtesy of Dr. Edgar Avalos}
  \label{fig:platon2}
\end{figure}

This study is motivated by two notable phenomena observed in the self-assembly of block copolymer (BCP) nanoparticles. First, despite their nanoscale dimensions--on the order of hundreds of nanometers--these particles consistently form well-defined polyhedral geometries, including cubes and octahedra, reminiscent of Platonic solids (see Fig.~\ref{fig:yabu}). Such geometric regularity at this scale is counterintuitive and challenges conventional expectations. Second, these structures are not rare anomalies; rather, they are reproducibly formed with high probability under specific, well-controlled experimental setups. These observations were first found in \cite{Avalos+2024} both in experiments and numerical simulations as in Fig.~\ref{fig:platon2}, and it was strongly suggested that the ratio of relaxation parameters of two unknowns in the model system (see Sec.2.2) is a key to have a consistent result with specific experimental setup. Nevertheless it remains a fundamental question: What underlying mathematical mechanisms produce the robust emergence of such exotic nanostructures and why such a ratio is important to detect desired morphologies? The goal of the paper is to answer these questions at least partially by using a simplified 1D model (6) for the polymer--solvent mixture.

Our model system has a free energy given by (5) based on the Ohta--Kawasaki type \cite{Ohta1986,OhtaIto1995} and its mathematical reformulation \cite{Nishiura1995} 
for the bulk system. As for the confined BCPs in 3D, extensive research has been conducted experimentally \cite{Yabu2014,Yan2018,Yan2023,Dai2020}. It has been demonstrated by a series of papers \cite{Avalos2016,Avalos2018,ashuraPaper} that the model system (5) reflects essential properties of nanoparticles in the solvent.

One of the fundamental questions is that how we reconstruct the global landscape of free energy, capturing all minimizers, saddle points, and transitional pathways. Moreover, how we can systematically control the system to a desired minimizer from a given initial condition that may have small fluctuation.

This is in general a formidable task, because the associated free energy is a rugged 
landscape that has so many minimizers and saddles in between. For instance, only 13 atoms in an appropriate potential field generate more than one thousand local minimizers (see \cite{Wales2003}).
Moreover, even if we fix all the parameters and the initial condition, the itinerancy of the trajectory and the final destination vary widely depending on the small fluctuation to the initial state, which renders the system dynamics very sensitive. 

On the other hand, in real chemical experiments or biological phenomena such as DNA folding, a targeted minimizer can be realized in a robust way without subtle external control, once the experimental setting is appropriately fixed. 
Nanoparticle formation of block copolymers is also a typical example in such a class presenting robustness and varieties of morphologies at the same time, and motivated us to study in this paper.
Note that there are two factors regarding to the experimental setting: One class consists of the polymer species, length of polymer chain, hydrophobicity or hydrophilicity, and other chemical properties that actually change the landscape of free energy. The other class is formed by pressure, temperature, initial concentration that don't affect the free energy, but may change the path of trajectories. We pay more attention on the second class of factors and clarify the relevancy to the dynamics of trajectories and the robustness of final morphologies against the external noise.

It would be worth noting here that when we trace the trajectory sliding down the energy landscape, we usually resort to steepest decent method, proceeding along the direction of the largest energy gradient. This is partly because the relaxation parameter for each variable is not a priori known and there is a common assumption (or belief) that free energy decreases at the fastest possible rate in nature. However, as noted in prior work \cite{Avalos+2024}, this approach often leads to a limited subset of minimizers that may diverge significantly from experimentally observed configurations. In fact if we have several unknowns, each variable could be accelerated or decelerated depending on the experimental setting such as pressure and concentrations. This suggests that speeding factors may be crucial for reproducing experimental conditions---such as pressure and initial concentration---and thus help to bridge the gap between experimental observations and theoretical predictions.\\

To achieve our goal, we propose two approaches that resolve these issues at least partially:
\begin{enumerate}

\item  {\bf{Atlas of saddle points.}}\\
  (a) {\it{Bifurcation path tracking}}: We explore all the saddle points of the free energy using continuation method. The saddle point search method, similar to bifurcation tracking, begins with approximate saddle solutions derived from trivial constant state, energy plateau configurations and good candidates obtained by the simplification method explained next. By systematically tracing solution branches along suitable parameters, this method allows us to comprehensively identify all the saddle points relevant to the concerning dynamics. Once these are identified, each minimizer can be easily obtained from saddles with low instability indices, enabling a global view of the free energy landscape.\\
  (b) {\it{Energy Landscape Simplification and Transformation}}:
  Certain adjustments of the parameters in the free energy can substantially simplify the landscape. These parameters depend on the specific problem; for example, in our problem, amplifying the hydrophilic-hydrophobic contrast restricts possible stable states (see the horizontal axis
  $\alpha$ in Fig.~\ref{bifAtlasTeX.eps}).
  By selecting appropriate parameters, one can simplify the energy landscape, facilitating the comprehensive identification of all minimizers. This approach yields valuable insights into the structure of the original, more complex landscape.

\item {\bf{Expansion and contraction of trajectory bundle for the multi-parameter family of PDEs.}}\\
  We consider multi-parameter family of PDEs parametrized by relaxation parameters. 
  A typical case is a one-parameter family of PDEs, parametrized by the ratio $r$ of relaxation parameters for a system with two variables. While steepest-descent dynamics are a common but suboptimal choice, this parameterization enables us to significantly broaden the accessible portion of the free energy landscape. Note that the relaxation parameters do not change the free energy landscape. What they change is the descending manner from high energy region to lower one, and if the ratio is far from 1, then the trajectory is almost parallel to the contour line of energy. 
  The initial state usually has high energy and quite unstable so that the trajectory behaves in a wild manner depending on the small random seed, and its final destination is not uniquely determined. To control such a behavior, the ratio $r$ plays a key role in the sense that it controls the expansion and contraction of bundle of trajectories parametrized by random seeds. Typically the contraction occurs when $r$ is far from 1 (see Fig.~\ref{fig:small_alpha}).

\end{enumerate}

\noindent
Combining these methods, we can find all the saddle points and minimizers relevant to our main issue, and how the trajectory itinerates a saddle network depending on the relaxation parameters before reaching the final destination. In fact, we can demonstrate numerically how a family of trajectories parametrized by relaxation parameter starting from the same random initial data sweeps out most of the saddle points and minimizers (see Fig.~\ref{fig:large_alpha} and Fig.~\ref{fig:small_alpha}).

As for the reproducibility of Platonic solids, the contraction of bundle of trajectories for $r\ll1, r\gg1$ and unique destination independent of random seeds support the experimental result that the desired morphologies can be obtained with high probability under noisy environment provided the process of precipitation (particle formation) is much faster than microphase separation. An intuitive reason of high reproducibility is that the trajectory circumvents dangerous region of saddle network and evolves along the specified path for $r\ll1$ and $r\gg1$ (see Sec.4).
In this paper we illustrate the efficiency of the approach by applying it to a simplified coupled Cahn--Hilliard (CCH) equations (6) in one-dimensional space.

The simplified model (6) can be obtained by setting $\sigma = 0$ for the original model (5), namely there are no chemical bonding between two homopolymers, nevertheless it keeps several key features such as rich structure of saddle network to which we can apply the above three methods. 

Here, we present several remarks on the significance of saddle points with high instability indices. The article \cite{Nishiura_Landscape2D} illustrates the critical role of high-index saddle points in the global free energy landscape for the 2D system (5), in which High--index Optimization--based Shrinking Dimer (HiOSD) method was used to explore the landscape of free energy. The HiOSD method is a powerful tool to detect the high-index saddles (see for instance \cite{P_Zhang_2019}), however, it does not tell us the detailed bifurcation structure of the saddle solutions as parameters are varied. Moreover, we are interested in more dynamic aspect of the CCH model such as the itinerancy of each trajectory and its parameter dependency on the ratio $r$. Nevertheless the integrated approach of the present and the HiOSD methods would become a powerful tool for higher dimensional problems.

Philosophically the view from highly unstable saddles is one of the keys to understand the complex dynamics	
such as collision dynamics among moving localized patterns \cite{Nishiura_scattering_2019,Nishiura_AMS_2009}. The trajectories are steered and sorted out by a network	
of saddle points and the Morse index is decreased along the trajectory.

The methodology presented here can be extended to the original system (5) with nonlocal term and even for higher dimensional spaces, which remains as a future work at present, however, the essence of our approach is clearly visible for the simplified model system without too much technicalities.

The contents of the paper is as follows. In Sec.2, we introduce the mathematical models and present their basic properties. Since the model satisfies an energy dissipation law with mass conservation, we use the structure preserving numerical scheme called the Discrete Variational Derivative Method (DVDM) to compute the trajectories.
In Sec.3, we present the global bifurcation diagram with respect to $\alpha$ (parameter related to affinity to the solvent), and clarify the network structure to understand the behavior of trajectory.
In Sec.4, based on the structure of saddle network in Sec.3, we try to understand the expansion and contraction of trajectory bundle with respect to the ratio $r$, which is a key observation to control the dynamics of trajectories and reduce the uncertainty coming from random seeds. Moreover, we study the itinerancy of each trajectory when the bundle is expanded based on the detailed saddle network structure in Sec.3. Finally, we conclude our discussion
and show the outlook and future problems in Sec.5.

\section{Model}

A typical example of an energy functional often considered in phase separation phenomena is the van der Waals-type functional given by
\begin{equation}
  F_{\epsilon_u} [u]
  = \int_\Omega \left\{ \frac{\epsilon_u^2}{2} |\nabla u|^2 + W(u) \right\} \dif x,
  \label{eqn:globalEnergy1}
\end{equation}
where $u$ is an order parameter, $\epsilon_u$ controls the size of the interface, and $\Omega$ is a smooth, bounded domain in $\mathbb{R}^N$.

In phase separation phenomena, at least two distinct phases can be observed, such as solid and liquid or burned and unburned regions.
In this study, we consider the two-phase system consisting of two homopolymers and the solvent.
Accordingly, for the potential function $W(u)$, we adopt the so-called double-well potential, which is typically expressed as a quartic polynomial:
\begin{equation}
  W_{\text{dw}}(u) = \frac{(1 - u^2)^2}{4}.
  \label{eqn:doublewell1}
\end{equation}
This potential attains its minimum at $u = \pm 1$, where we interpret $u = - 1$ as the solvent-rich phase and $u = 1$ as the polymer-rich phase.
In other words, the order parameter $u$ represents the shape of polymer-aggregated particles.

Furthermore, we assume that the polymer is a diblock copolymer, meaning that it consists of two chemically distinct polymer components, A and B, joined via copolymerization.
The energy functional associated with this diblock copolymer system is defined as
\begin{equation}
  F_{\epsilon_v ,\sigma }[v] =\int\limits_{\Omega }\left\{ \frac{\epsilon _{v}^{2}}{2}\left\vert \nabla v\right\vert ^{2}+W\left(v\right) +\frac{\sigma }{2}\left\vert \left( -\Delta \right)
  ^{-1/2}\left( v-\overline{v}\right) \right\vert ^{2}\right\} \dif x.
  \label{eqn:globalEnergy2}
\end{equation}
Here, $\overline{v}$ represents the spatial average, i.e. $\overline{v}:=\int_{\Omega} v \dif x/(\int_{\Omega} \dif x)$, and the potential function is again taken to be the double-well potential:
\begin{equation}
  W_{\text{dw}}(v) = \frac{(1 - v^2)^2}{4}.
  \label{eqn:doublewell2}
\end{equation}
In this case, the order parameter $v$ characterizes the microphase separation within the particles.
The regions where $v = -1$ and $1$ correspond to polymer A-rich and polymer B-rich domains, respectively. The energy (\ref{eqn:globalEnergy2}) was coined by \cite{Ohta1986} and reformulated mathematically by \cite{Nishiura1995}.

\subsection{Free energy and the coupled Cahn--Hilliard equations}

In the binary polymer mixture considered here, the two order parameters, $u$ and $v$, evolve over time to minimize the following energy functional (see \cite{OhtaIto1995, Avalos2016}):
\begin{subequations}\label{eqn:energy}
  \begin{gather}
    F[u,v](t) = \int_{\Omega} G[u,v](t,x) \dif x,
    \label{eqn:globalEnergy3}
    \\
    G[u,v] = \frac{\epsilon_u^2}{2}|\nabla u|^2 + \frac{\epsilon_v^2}{2}|\nabla v|^2 + W(u,v) + \frac{\sigma}{2}\left| (-\Delta)^{-1/2}(v-\bar{v}) \right|^2
    \label{eqn:functional3}
    \\
    W(u,v) = \frac{(1-u^2)^2}{4} + \frac{(1-v^2)^2}{4} + \alpha uv + \beta uv^2.
  \end{gather}
\end{subequations}
Broadly speaking, the level set $u = 0$ represents the boundary of the polymer particles.
The coefficients $\epsilon_u$ and $\epsilon_v$ of the gradient energy term are parameters that define the interface thickness of the phase separation.
The parameter $\alpha$ controls the interfacial affinity of polymers A and B for the solvent:
Depending on its sign, it indicates which polymer is more soluble.
For instance, if $\alpha > 0$, the energy is minimized when $u$ and $v$ have opposite signs.
Given that $u \sim -1$ corresponds to the solvent-rich phase, this implies that polymer B is more soluble than polymer A.
The term $\beta < 0$ ensures that $u \sim 1$ in regions where polymers are present $(v^2 > 0)$, thereby indicating that $u$ represents the polymer domain. Note that the well-posedness for the Euler--Lagrange equations associated with the energy (\ref{eqn:energy}) has been proved by \cite{CherfilsMiranville2022,PrimioGrasselli2022}.
In this paper we focus on the following simplified problem (\ref{eqn:CCH}) in one-dimensional spatial domain with $\Omega := [0,L)$. We also assume $\sigma = 0$, corresponding to a scenario in which the two polymers are not covalently bonded.
  The following periodic boundary condisions are imposed on the order parameters $u, v$ and the variational derivatives of the local free energy $G$:
  \begin{gather*}
    u(t,0) = u(t,L), \quad
    \left.\frac{\partial u(t,x)}{\partial x}\right|_{x=0} = \left.\frac{\partial u(t,x)}{\partial x}\right|_{x=L},
    \\
    v(t,0) = v(t,L), \quad
    \left.\frac{\partial v(t,x)}{\partial x}\right|_{x=0} = \left.\frac{\partial v(t,x)}{\partial x}\right|_{x=L},
    \\
    \frac{\delta G[u,v]}{\delta u}(t,0) = \frac{\delta G[u,v]}{\delta u}(t,L), \quad
    \left.\frac{\partial}{\partial x}\frac{\delta G[u,v]}{\delta u}(t,x)\right|_{x=0} = \left.\frac{\partial}{\partial x}\frac{\delta G[u,v]}{\delta u}\right|_{x=L},
    \\
    \frac{\delta G[u,v]}{\delta v}(t,0) = \frac{\delta G[u,v]}{\delta v}(t,L), \quad
    \left.\frac{\partial}{\partial x}\frac{\delta G[u,v]}{\delta v}\right|_{x=0} = \left.\frac{\partial}{\partial x}\frac{\delta G[u,v]}{\delta v}\right|_{x=L},
  \end{gather*}
  for all $t > 0$.

  Using the variational derivatives of the local free energy $G$ with $\sigma=0$, we derive the following coupled Cahn--Hilliard (CCH) system:
  \begin{equation}
    \left\{
    \begin{aligned}
      & \tau_u \frac{\partial u(t,x)}{\partial t} = \frac{\partial^2}{\partial x^2} \frac{\delta G[u,v]}{\delta u}
      = \frac{\partial^2}{\partial x^2}\Big[ -\epsilon_u^2 \frac{\partial^2}{\partial x^2} u - u + u^3 + \alpha v + \beta v^2 \Big]
      \ \text{for}\ t>0, x\in[0,L),
        \\
        & \tau_v \frac{\partial v(t,x)}{\partial t} = \frac{\partial^2}{\partial x^2} \frac{\delta G[u,v]}{\delta v}
        = \frac{\partial^2}{\partial x^2}\Big[ -\epsilon_v^2 \frac{\partial^2}{\partial x^2} v - v + v^3 + \alpha u + 2\beta uv \Big]
        \ \text{for}\ t>0, x\in[0,L).
    \end{aligned}
    \right.
    \label{eqn:CCH}
  \end{equation}
  \noindent
  The system (\ref{eqn:CCH}) exhibits a rich structure and retains several key features of (\ref{eqn:energy}), as will be demonstrated in the following sections.

  \subsection{Relaxation parameters}

  In the CCH system, the parameters $\tau_u$, $\tau_v > 0$ represent the relaxation parameters that control the temporal evolution of $u$ and $v$, respectively.
  By changing the time scale, we can reformulate the system into an equivalent system where $\tau_u$ and $\tau_v$ are replaced by $r := \tau_u / \tau_v$ and $1$, respectively. Here $r$ is the ratio of two relaxation parameters that plays a key role in the subsequent sections. When $r$ becomes small, the relaxation of $u$ is faster than that of $v$; when $r$ becomes large, vice versa. 
  Avalos et al. \cite{Avalos+2024} numerically simulated the CCH system (5) and successfully reproduced the formation of polyhedral structures of block copolymers.
  The difference in relaxation parameters in the system corresponds to variations in solvent evaporation rates or pressure in experimental settings.
  In a standard gradient-flow system, they are typically set to $1:1$, leading to a steepest-descent dynamics in the energy landscape.
  However, in this study, we investigate the system's dynamics by perturbing the ratio away from unity, which corresponds to an oblique descent in the energy landscape. Note that the resulting system with different relaxation parameters is no longer a gradient system of the free energy.
  This approach enables us to numerically explore local minima and saddle points that would be difficult to reach via steepest-descent dynamics.
  Through this analysis, we aim to elucidate the free energy landscape of the CCH system.

  \subsection{Some properties of the coupled Cahn--Hilliard system}

  The CCH system satisfies two fundamental properties: The conservation law and the dissipation law.
  By integrating the first equation in \eqref{eqn:CCH} over the domain $[0,L)$ and using the boundary conditions for the variational derivative of the local energy, we obtain
    \begin{align}
      \frac{\dif}{\dif t}\int_0^L u(t,x) \dif x = 0.
    \end{align}
    This implies that if the initial condition is given by $u(t=0, x) = u_0(x)$, then for any $t > 0$,
    \begin{align}
      \bar{u} = \frac{1}{L} \int_0^L u(t,x) \dif x = \frac{1}{L} \int_0^L u_0(x) \dif x.
    \end{align}
    A similar argument holds for $v$, ensuring that if $v(t=0, x) = v_0(x)$, then 
    \begin{align}
      \bar{v} = \frac{1}{L} \int_0^L v(t,x) \dif x = \frac{1}{L} \int_0^L v_0(x) \dif x
    \end{align}
    for all $t > 0$.

    The energy dissipation law can be derived by differentiating the total energy functional \eqref{eqn:globalEnergy3} with respect to time.
    Using the periodic boundary conditions for $u$, $v$, $\frac{\delta G}{\delta u}$, and $\frac{\delta G}{\delta v}$, we obtain from \eqref{eqn:CCH}:
    \begin{align}
      \frac{\dif}{\dif t} F[u,v](t) &= \int_0^L \left(  \frac{\partial u}{\partial t}\frac{\delta G[u,v]}{\delta u} + \frac{\partial v}{\partial t}\frac{\delta G[u,v]}{\delta v} \right)  \dif x
      \nonumber\\
      &= - \int_0^L \left( \left| \frac{\partial}{\partial x}\frac{\delta G[u,v]}{\delta u} \right|^2 + \left| \frac{\partial}{\partial x}\frac{\delta G[u,v]}{\delta v} \right|^2 \right)  \dif x
      \leq 0.
      \label{eqn:dissipation}
    \end{align}
    This represents the dissipation of energy over time.

    To analyze the stability of the system around the homogeneous state $(u,v) = (0,0)$, we consider the linearized equations:
    \begin{equation}
      \left\{
      \begin{aligned}
        & \frac{\partial u(t,x)}{\partial t} = \frac{1}{\tau_u} \frac{\partial^2}{\partial x^2}\Big[ -\epsilon_u^2 \frac{\partial^2}{\partial x^2} u - u + \alpha v \Big]
        \ \text{for}\ t>0, x\in[0,L),
          \\
          & \frac{\partial v(t,x)}{\partial t} = \frac{1}{\tau_v} \frac{\partial^2}{\partial x^2}\Big[ -\epsilon_v^2 \frac{\partial^2}{\partial x^2} v - v + \alpha u \Big]
          \ \text{for}\ t>0, x\in[0,L).
      \end{aligned}
      \right.
      \label{eqn:LCCH}
    \end{equation}
    Assuming plane-wave solutions of the form
    \begin{equation}
      \largevec{u}{v} = \sum_{n\in\mathbb{N}} \ope^{\lambda t} \ope^{\opi \omega_n x} \largevec{u_n}{v_n},
    \end{equation}
    and substituting into \eqref{eqn:LCCH}, we obtain the eigenvalue problem
    \begin{equation}
      \largevec{0}{0} = 
      \largemat{
        \frac{1}{\tau_u}\left[ -\epsilon_u^2 \omega_n^4 + \omega_n^2 \right] - \lambda
      }{
        -\frac{1}{\tau_u}\alpha\omega_n^2
      }{
        -\frac{1}{\tau_v}\alpha\omega_n^2
      }{
        \frac{1}{\tau_v}\left[ -\epsilon_v^2\omega_n^4 + \omega_n^2 \right] - \lambda
      }
      \largevec{u_n}{v_n},
    \end{equation}
    where $\omega_n := \frac{2\pi n}{L}$.
    Since the determinant of the coefficient matrix must be zero, we obtain the characteristic equation:
    \begin{equation}
      \left( \lambda - \frac{1}{\tau_u}\left[ -\epsilon_u^2\omega_n^4 + \omega_n^2 \right] \right)
      \left( \lambda - \frac{1}{\tau_v}\left[ -\epsilon_v^2\omega_n^4 + \omega_n^2 \right] \right)
      - \frac{1}{\tau_u\tau_v}\alpha^2\omega_n^4
      = 0.
    \end{equation}

\begin{figure}
      \hfil\includegraphics[scale=.5]{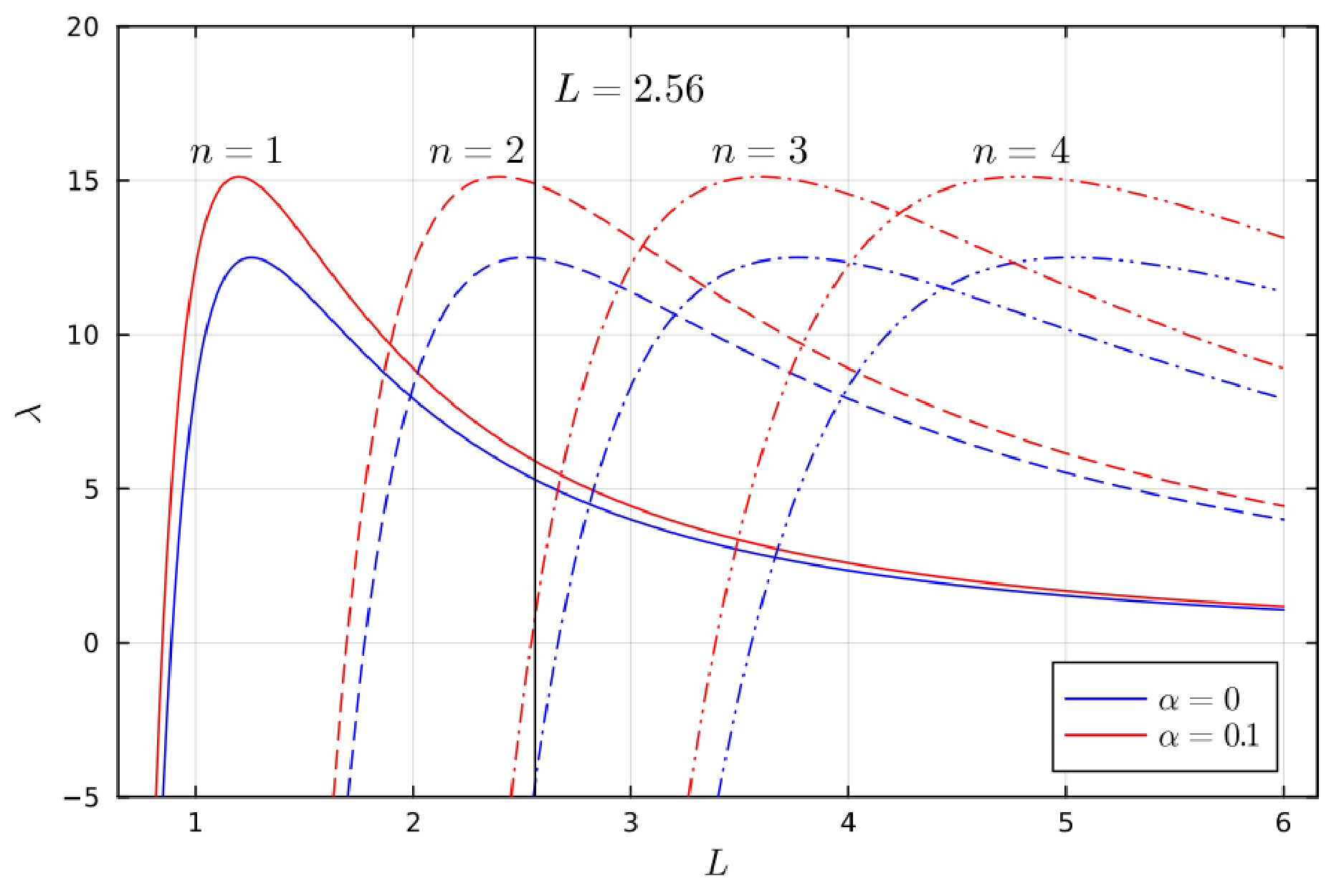}
      \caption{
        Dispersion relations when $\alpha=0$ (blue line) or $\alpha=0.1$ (red line) with $\tau_u=\tau_v=1$.
        In the setting $L=2.56$ of this paper, the second mode exhibits the highest instability and is the first to grow.
      }
      \label{fig:disp_rel_a0_a01}
    \end{figure}

    For $\alpha = 0$, the dispersion relations reduce to those of the standard Cahn--Hilliard equation:
    \begin{align}
      \lambda = \frac{\omega_n^2}{\tau_u}\left( -\epsilon_u^2\omega_n^2 + 1 \right), \ \frac{\omega_n^2}{\tau_v}\left( -\epsilon_v^2\omega_n^2 + 1 \right)
    \end{align}
    In this study, we fix $\epsilon_u^2 = \epsilon_v^2 = 0.02$, which corresponds to a characteristic wavelength of $2\pi / \sqrt{1/2\epsilon_u^2} \sim 1.26$ and $L=2.56$.
    As shown by the blue curves in Fig.~\ref{fig:disp_rel_a0_a01}, the most unstable wave number is $n=2$.
    Since the domain length is $L = 2.56$, the initial phase separation at $\alpha = 0$ results in the growth of two wavelengths.
    For $\alpha \neq 0$, assuming $\epsilon := \epsilon_u = \epsilon_v$, the dispersion relation takes the form
    \begin{align}
      \lambda = \frac{1}{2}\left( \frac{1}{\tau_u} + \frac{1}{\tau_v} \right)(-\epsilon^2\omega_n^4 + \omega_n^2) 
      \pm \sqrt{\left[\frac{1}{2}\left( \frac{1}{\tau_u} - \frac{1}{\tau_v} \right)(-\epsilon^2\omega_n^4 + \omega_n^2)\right]^2 + \frac{\alpha^2\omega_n^2}{\tau_u\tau_v}}.
    \end{align}
    For $\tau_v = \tau_u = 1$, this simplifies to
    \begin{align}
      \lambda = \omega_n^2(-\epsilon^2\omega_n^2 + 1 \pm \alpha),
    \end{align}
    which is represented by the red curves in Fig.~\ref{fig:disp_rel_a0_a01}.
    As $|\alpha|$ increases, the most unstable wave number also increases.

    In the case where $\tau_v \gg \tau_u = 1$ ($r \ll 1$), the phase separation of $v$ occurs significantly more slowly than that of $u$.
    The dispersion relation for $u$ approximates that of the single Cahn--Hilliard equation:
    \begin{align}
      \lambda \sim \frac{\omega_n^2}{\tau_v}(-\epsilon^2\omega_n^2 + 1),
    \end{align}
    which also holds in the opposite limit, $\tau_u \gg \tau_v = 1$ ($r \gg 1$).

    \subsection{Structure preserving scheme}

    We introduce a discrete numerical scheme based on the Discrete Variational Derivative Method (DVDM) developed by Furihata et al. \cite{DVDM}, 
    which is known as a stable numerical scheme effective for dissipative systems, including the Cahn--Hilliard equation.
    We define the discrete local energy $\Gd$ corresponding to the functional \eqref{eqn:functional3} as follows:
    \begin{align}
      \Gd(\vecU, \vecV) =
      &\frac{\epsilon_u^2}{2} \left( \frac{\dkp\Uk + \dkm\Uk}{2} \right)^2
      + \frac{\epsilon_v^2}{2} \left( \frac{\dkp\Vk + \dkm\Vk}{2} \right)^2
      \nonumber\\
      & + \frac{\left( 1 - \Uk^2 \right)^2}{4} + \frac{\left( 1 - \Vk^2 \right)^2}{4}
      + \alpha\Uk\Vk - \beta\Uk\Vk^2,
    \end{align}
    where $\vecU = (U_1,\ldots,U_K)^{\mathrm{T}}$ and $\vecV = (V_1,\ldots,V_K)^{\mathrm{T}}$.
    By computing the discrete variational derivative of this discretized local free energy, we obtain the following discrete numerical scheme:
    \begin{equation}\tag{DCCH}
      \left\{
      \begin{aligned}
        & \tau_u \frac{\UU{k}{m+1} - \UU{k}{m}}{\Delta t} = \dktwo \left( \frac{\delta\Gd}{\delta(\vecU[m+1], \vecU[m])} \right)_k,
        \\
        & \left( \frac{\delta\Gd}{\delta(\vecU[m+1], \vecU[m])} \right)_k = 
        -\epsilon_u^2 \dktwo \frac{\UU{k}{m}+\UU{k}{m+1}}{2} - \frac{\UU{k}{m}+\UU{k}{m+1}}{2}
        \\
        &\qquad + \frac{1}{4}\left( \bigpw{\UU{k}{m}}{3} + \bigpw{\UU{k}{m}}{2}\UU{k}{m+1} + \UU{k}{m}\bigpw{\UU{k}{m+1}}{2} + \bigpw{\UU{k}{m+1}}{3} \right)
        \\
        &\qquad + \alpha \frac{\VV{k}{m}+\VV{k}{m+1}}{2} + \beta \frac{\bigpw{\VV{k}{m}}{2}+\bigpw{\VV{k}{m+1}}{2}}{2},
        \\
        & \tau_v \frac{\VV{k}{m+1} - \VV{k}{m}}{\Delta t} = \dktwo \left( \frac{\delta\Gd}{\delta(\vecV[m+1], \vecV[m])} \right)_k,
        \\
        & \left( \frac{\delta\Gd}{\delta(\vecV[m+1], \vecV[m])} \right)_k = 
        -\epsilon_v^2 \dktwo \frac{\VV{k}{m}+\VV{k}{m+1}}{2} - \frac{\VV{k}{m}+\VV{k}{m+1}}{2}
        \\
        &\qquad + \frac{1}{4}\left( \bigpw{\VV{k}{m}}{3} + \bigpw{\VV{k}{m}}{2}\VV{k}{m+1} + \VV{k}{m}\bigpw{\VV{k}{m+1}}{2} + \bigpw{\VV{k}{m+1}}{3} \right)
        \\
        &\qquad + \alpha \frac{\UU{k}{m}+\UU{k}{m+1}}{2} + \beta \frac{(\UU{k}{m}+\UU{k}{m+1})(\VV{k}{m}+\VV{k}{m+1})}{2},
      \end{aligned}
      \right.
    \end{equation}
    Here, $\UU{k}{m}$ and $\VV{k}{m}$ are the discrete approximations of $u(t,x)$ and $v(t,x)$ at $(t,x) = (m \Delta t, k \Delta x)$, respectively.
    The numerical parameters are set as follows: The time step size is $\Delta t = 10^{-4}$; the spatial grid size is $\Delta x = 0.02$; the number of grid points is $K = 128$; and the domain length is given by $L = K \Delta x = 2.56$.
    Additionally, we fix the coefficients for the gradient energy terms as $\epsilon_u^2 = \epsilon_v^2 = 0.02$, which corresponds to an unstable wavelength of approximately $L/2$.
    The second interaction term coefficient is fixed as $\beta = -0.3$.

\section{Global bifurcation diagram with respect to $\alpha$}
\begin{figure}
  \hfil\includegraphics[width=\hsize]{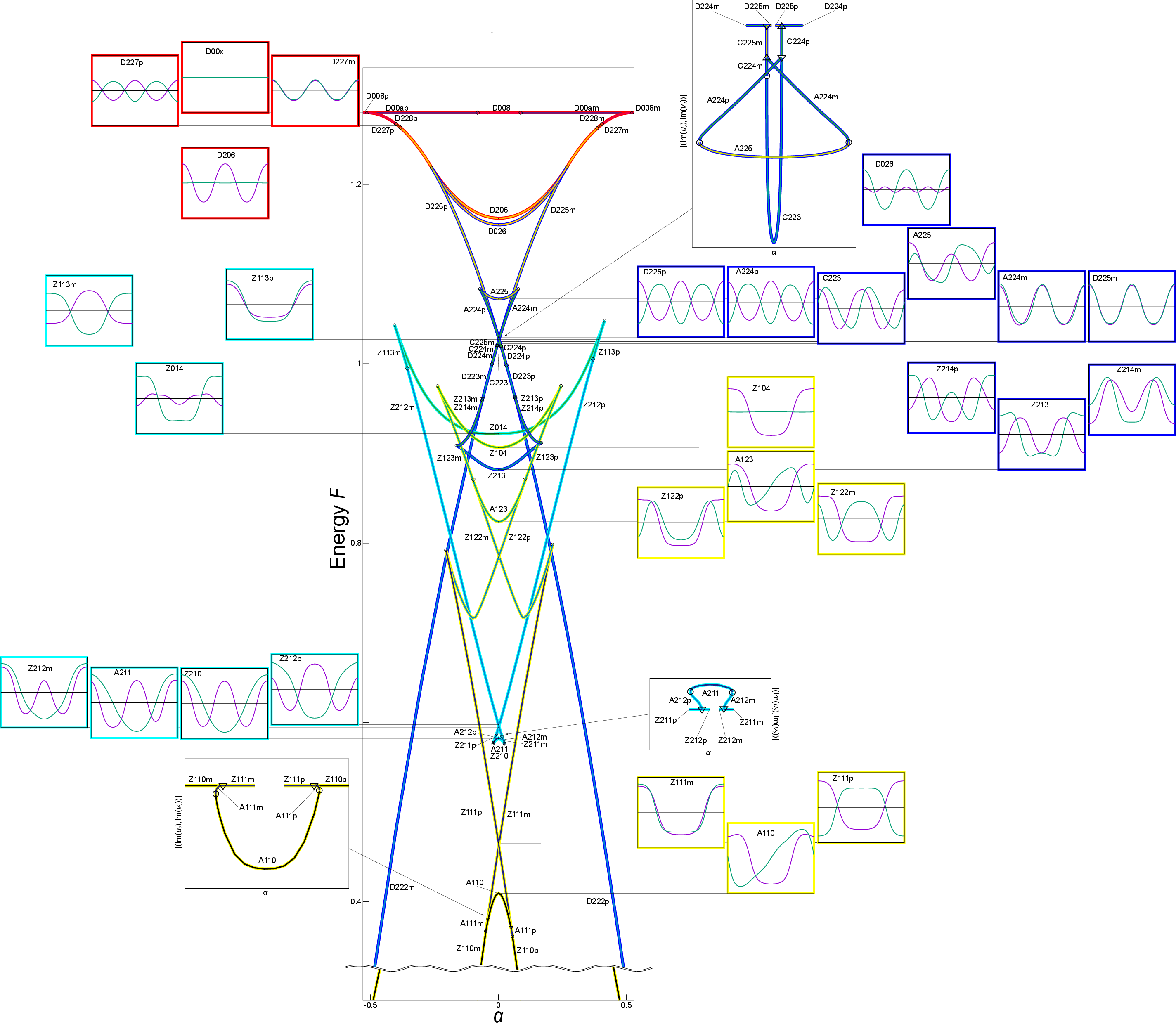}
  \caption{Global bifurcation diagram and corresponding patterns.
    Branches are shown in combinations of two lines: Background thick line
    (base line) and overdrawn  thin line (center line).
    Relationship between the line combination and the type of the solution
    is shown in Fig.~\ref{bifAtlasTeXLegend.eps}.
    Four colors of base lines, red, blue, yellow, and cyan, corresponds
    to four connected components:
    ``linear regime component'',
    ``two-hump component'',
    ``one-mode microphase component'', and
    ``one-hump component'', respectively.
    Colors of center lines indicate Morse index.
    Details of each connected component is shown in Figs.
    \ref{bifAtlasTeX1.eps},
    \ref{bifAtlasTeX2.eps},
    \ref{bifAtlasTeX3.eps}, and
    \ref{bifAtlasTeX4.eps}, respectively.
    Bifurcation points are shown by different symbols
    as shown in Fig.~\ref{bifAtlasTeXLegend.eps}.
  }\label{bifAtlasTeX.eps}
\end{figure}

\begin{figure}
  \hfil\includegraphics[width=\hsize]{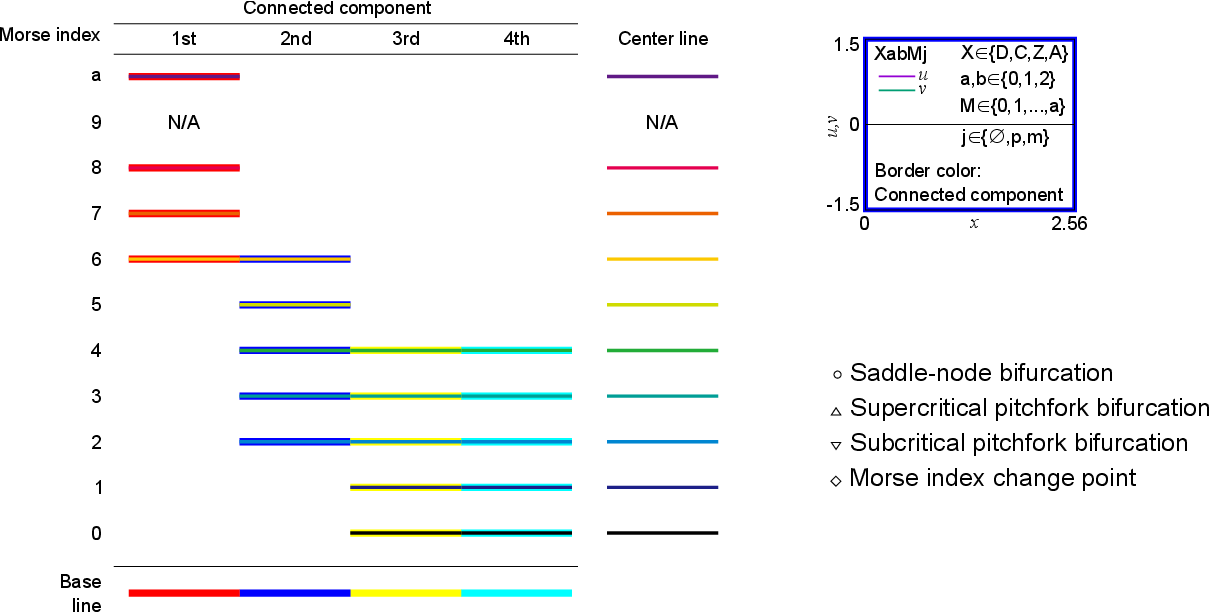}
  \caption{Legend of Fig.~\ref{bifAtlasTeX.eps}.
  }\label{bifAtlasTeXLegend.eps}
\end{figure}

Global bifurcation diagram is shown in Fig.~\ref{bifAtlasTeX.eps} and the Legend is shown in Fig.~\ref{bifAtlasTeXLegend.eps}. 

\noindent
Information of global bifurcation structure gives an atlas for
the exploration of the landscape because stable and unstable
manifolds of each saddle and their connections play essential role
for the asymptotic state of the system.
$\alpha$ is chosen for the bifurcation parameter.
Because $\alpha$ represents the strength of separation of two polymers,
two polymers easily separated each other when $|\alpha|\gg0$
and therefore emerging patterns tend to become simple for larger $|\alpha|$.
As the result many patterns appear for small $|\alpha|$ whereas
only a small number of patterns appear for large $|\alpha|$
as shown in Fig.~\ref{bifAtlasTeX.eps}.
Thus, global bifurcation structure is investigated in detail for
the range $-0.53\le\alpha\le0.53$. There are three local minimizers; ${\tt A110}$ called the Janus particle (JP), ${\tt Z210}$ the Single particle (SP), and ${\tt Z110j}$. See Sec.3.2 for the notation.
The location and the interrelation among them are discussed in Sec.3.3.

\subsection{Numerical method}
Computation of the global bifurcation structure and saddle network requires
significantly greater computational effort compared to
direct time integration.
If we were to apply the structure preserving
scheme---which guarantees mass conservation and monotonic decrease of
energy---as is,
the computation would take several decades to complete.
Therefore, in this section, we employ a spectral method to reduce
computational time.
The following points should be noted:
\begin{itemize}
\item
  Mass is conserved exactly, up to numerical round-off errors.
\item
  Monotonic energy decay is generally preserved;
  in our simulations, no energy increase has been observed.
\item
  For several critical computations, both schemes have been implemented
  and cross-validated.
\end{itemize}
An important side benefit of using spectral methods is that it
facilitates the analysis of solution symmetries.
For instance, if a solution is stable within a symmetric subspace
but unstable under symmetry-breaking perturbations,
computations and stability analysis restricted to the symmetric
subspace become more tractable.
Moreover, by restricting the set of Fourier modes, it is possible
to selectively enforce certain symmetries on the solutions.

The functions $u$ and $v$ are expanded as discrete Fourier series,
\begin{equation}
  (u(x,t),v(x,t))
  =\sum_{{n}=-\left[\frac{{N}}{2}\right]}^{\left[\frac{{N}-1}{2}\right]}
  \ope^{\opi \frac{2\oppi}{L}{n} x}(u_{n}(t),v_{n}(t)),\;\;\;L=2.56.
  \label{eqn:SpectralMethod}
\end{equation}
For consistency with the finite difference computations, we set $N=127$.

Since the nonlinear terms are at most cubic, we apply ``four halves method'',
which is an extension of ``three halves method'',
to eliminate aliasing errors.
First we apply the inverse Fourier transform using $2N+1$ modes
adding $u_n=v_n=0$ for $N/2+1\le |n|\le N$.
Then compute the nonlinearities in physical space, and perform a Fourier
transform.
Because the maximum order of the Fourier mode is $3N/2$,
aliasing errors only appear in $[N/2]+1\le |n|\le N$.
Therefore we can remove the aliasing errors truncating them.
As the result of this procedure, the original degree of freedom
$N$, $-[N/2]\le n\le [(N-1)/2]$, is restored without contamination.

Equation (\ref{eqn:SpectralMethod}) is substituted into Eq. (\ref{eqn:CCH})
and then equations can be simply described by
\[
\frac{\opd\vec z}{\opd t}=\vec f(\vec z),
\]
where $\vec z=\{u_n,v_n\}$.
Phase-space vector $\vec z$ is integrated numerically using the forward Euler
method together with the Crank-Nicolson method.

Steady solutions of $\vec z$ satisfies the condition $\vec z(t')=\vec z(0)$
for an arbitrary $t'>0$.
This condition is approximated numerically as
\[
\left(\vec z(0.1)-\vec z(0)\right)\overline{\left(\vec z(0.1)-\vec z(0)\right)}
<1\times10^{-10},
\]
where $\overline{\cdot}$ denotes the conjugate pair.
Because generic initial value $\vec z(0)$ does not satisfy the above
condition, the Newton-Raphson method is applied together with the
GMRes method \cite{Watanabeetal2010} to obtain approximated steady
solution $\vec z(0)$.
Further, the Arnoldi method is utilized to solve eigenvalue problems
to examine the linear stability of the solutions.

\subsection{Symmetry and classification of solutions}
Spectral method is appropriate for the discussion of symmetry because
the solution is naturally extracted into the spectrum.
The solution which has the highest symmetry is constant solution.
This is a dihedral group $D_\infty$, which has reflection symmetry
and infinite rotational symmetry.
Constant solution is always unstable and in the range
$-0.53\le\alpha\le0.53$, the most unstable mode is $2$ in the current
system size ($L=2.56$)
as shown in Fig.~\ref{fig:disp_rel_a0_a01}. 
First symmetry breaking from $D_\infty$ occurs with the emergence of
dihedral group $D_2$ via pitchfork bifurcation.
The typical solutions belonging to this group are
${\tt D227p}$ and ${\tt D227m}$.

There are two different types of symmetry breaking for $D_2$.
One only loses the $180^\circ$ rotational symmetry
whereas the other only loses the reflection symmetry.
The former is $C_2$, the cyclic group of order 2, whereas the
latter, $Z_2:=D_2/C_2$.
From the perspective of the Fourier coefficients,
the $C_2$ consists of complex coefficients $u_{2n}$ and $v_{2n}$,
whereas $Z_2$ consists of real coefficients $u_n$ and $v_n$.
In fact, $C_2\cong Z_2$, however these are treated as the
different group so as to distinguish the nature of the solution.
When $C_2$ or $Z_2$ loses symmetry, any solutions which has no symmetry
are represented by $A$.

In sum, any solutions are classified into
$D_\infty$, $D_2$, $C_2$, $Z_2$, and $A$.
Hereafter we use the symbols $D$, $C$, $Z$, and $A$ for the sake of
simplicity because the order of $C$ and $Z$ are 2,
and the order of $D$ is 2 except for $D_\infty$ for the constant solution.

Using these symbols, solutions are expressed as
\[
XabMj,\;
X\in\{{\tt D},{\tt C},{\tt Z},{\tt A}\},\;
a,b\in\{{\tt 0},{\tt 1},{\tt 2}\},\;
M\in\{{\tt 0},{\tt 1},\cdots,{\tt 9},{\tt a},\cdots,{\tt f}\},\;
j\in\{\emptyset,{\tt p},{\tt m}\}.
\]
$X$ represents the symmetry of the solution,
$a$ represents the dominant mode of $u$,
$b$ represents the dominant mode of $v$,
$M$ represents the Morse index appearing in the sequence of
hexadecimal digits,
and $j$ represents the sign of $u$ and $v$, respectively.
Eq. (\ref{eqn:CCH}) shows that $(u,-v,-\alpha)$ is also the solution
if $(u,v,\alpha)$ is the solution when $\bar u=\bar v=0$.
In the present study, although $\bar u,\bar v\neq0$, this relationship
is approximately satisfied because they are small enough.
As the result, $j=\emptyset$ for branches which is approximately symmetric
with respect to $\alpha=0$.
Branch which does not have this symmetry has corresponding pair
with the opposite sign.
In this case, if one is $j={\tt p}$, then the other is $j={\tt m}$
and vise versa.
Almost all branches have corresponding pair each other, however,
there is only one exception due to the non-zero values of
$\bar u$ and $\bar v$
(${\tt C225m}$, see the inset of two-hump component
in Fig.~\ref{bifAtlasTeX2.eps}).
Generally $a$ and $b$ are indices at which $|u_{a}|$ and $|v_{b}|$ reach
their maximum values, respectively.
However, in some branches $a$ or $b$ not always gives maximum value
along the branch.
In such cases, a special notice will be provided.

\subsection{Details of bifurcation structure}
Although bifurcation structure shown in Fig.~\ref{bifAtlasTeX.eps}
is complicated, it can be clearly understood from the viewpoint
of connected components.
It consists of four connected components, the first, second, third,
and fourth connected component in descending order of the maximum energy,
which are shown in Figs.
\begin{figure}
  \hfil\includegraphics[width=\hsize]{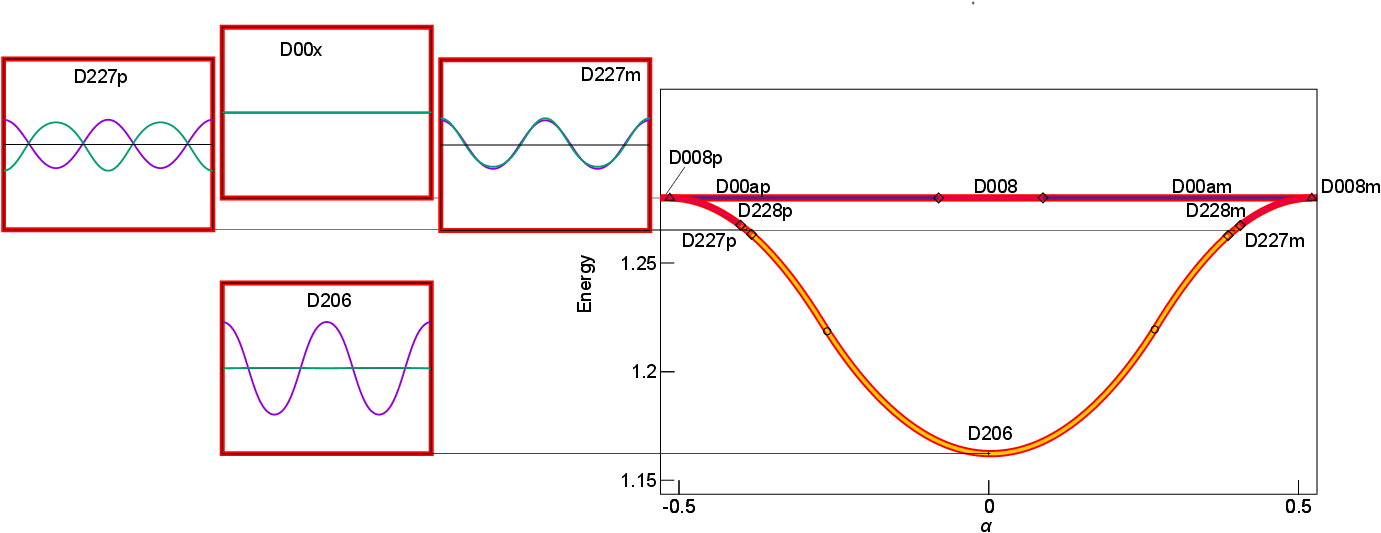}
  \caption{The first connected component,
    linear regime component.
    This component consists of constant solution and first
    bifurcation from it.
    $\hat u\sim\hat v$ emerges at $\alpha\sim0.5$, and
    $\hat u\sim-\hat v$ emerges at $\alpha\sim-0.5$,
    and the both are connected by the symmetric branch ${\tt D206}$.
    Note that $\hat u$ and $\hat v$ are eigenfunctions
    (see Sec.~\ref{Sec:Saddlenetwork}).
  }\label{bifAtlasTeX1.eps}
\end{figure}
\ref{bifAtlasTeX1.eps},
\begin{figure}
  \hfil\includegraphics[width=\hsize]{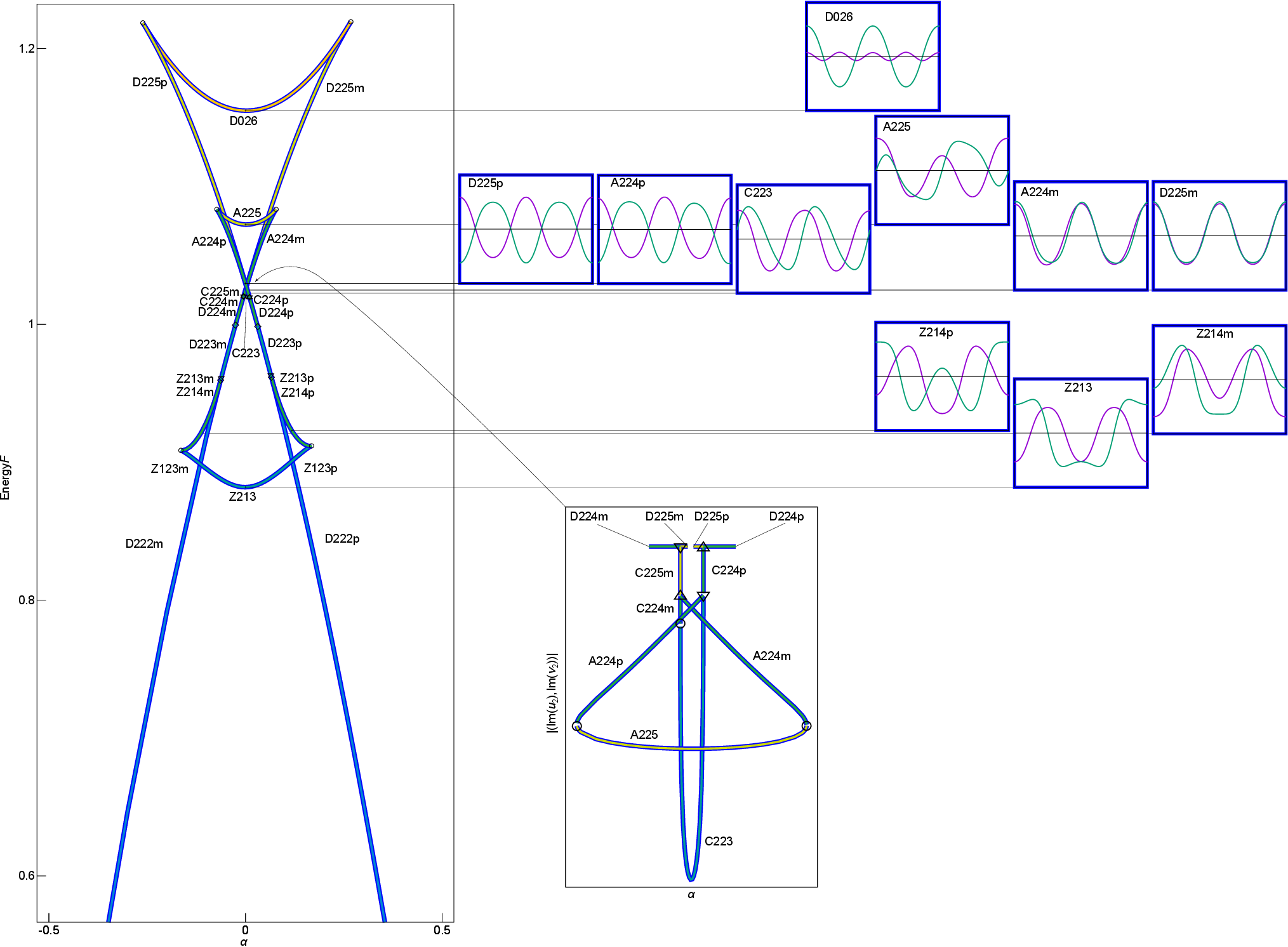}
  \caption{The second connected component,
    two-hump component.
    Fourier coefficients $u_{\pm2}$ is dominant in this connected component.
    Topology of the bifurcation structure is symmetric with respect
    to $\alpha=0$, except for the bifurcation of ${\tt C225m}$.
    At the bifurcation point ${\tt D224m}$ to ${\tt D225m}$,
    ${\tt C225m}$ emerges via subcritical pitchfork bifurcation whereas
    at the bifurcation point ${\tt D224p}$ to ${\tt D225p}$,
    ${\tt C224p}$ emerges via supercritical pitchfork bifurcation.
    At these bifurcation points solutions lose reflection symmetry.
    Further symmetry breaking occurs at pitchfork bifurcation points of
    ${\tt A224m}$ and ${\tt A224p}$.
    The former is supercritical and the latter subcritical as the
    result of asymmetry of the topological bifurcation structure.
    Another symmetry breaking occurs via subcritical pitchfork bifurcation
    at the point ${\tt D222}j$ to ${\tt D223}j$.
  }\label{bifAtlasTeX2.eps}
\end{figure}
\ref{bifAtlasTeX2.eps},
\begin{figure}
  \hfil\includegraphics[width=\hsize]{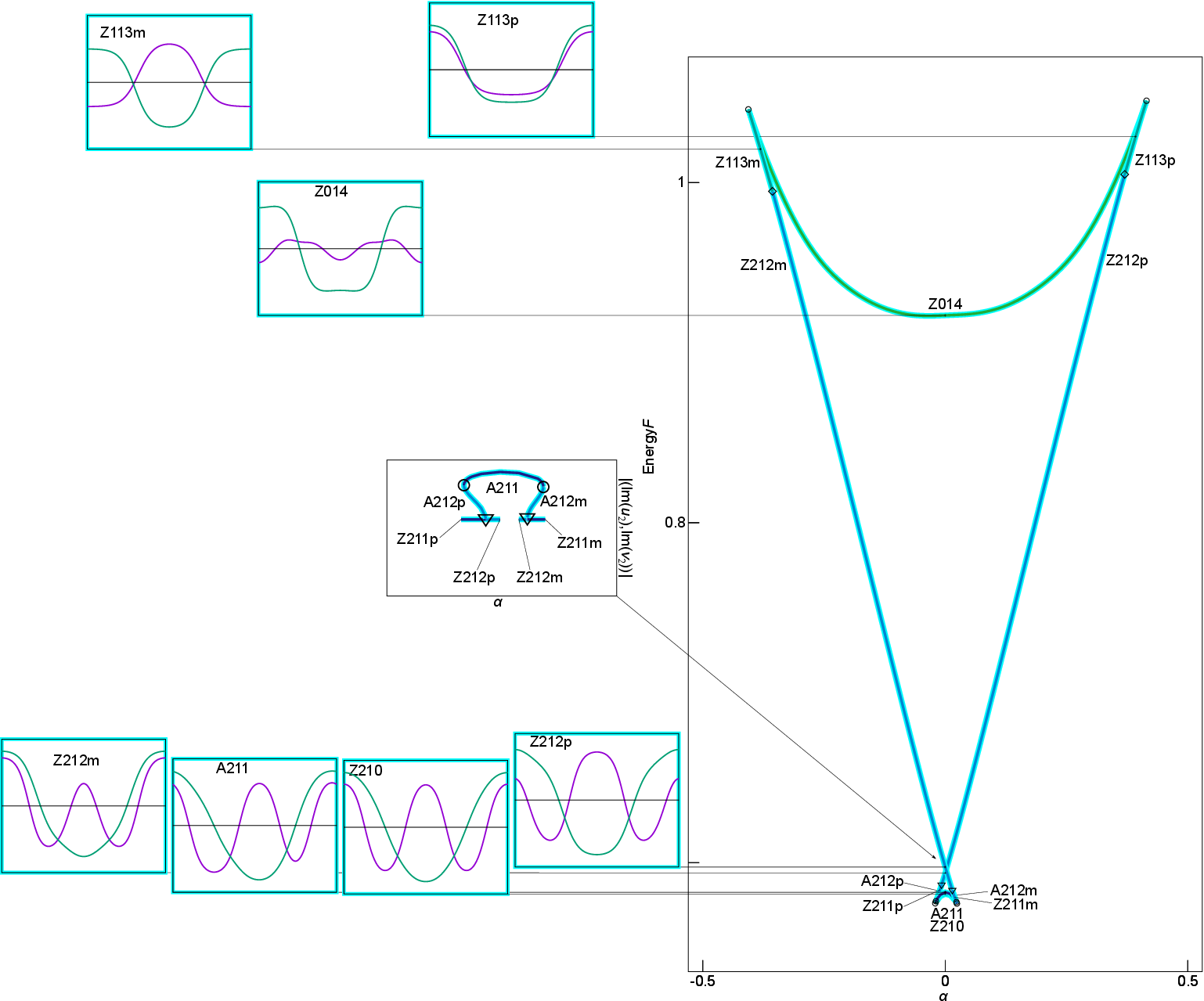}
  \caption{The third connected component,
    one-mode microphase component.
    Fourier coefficients $v_{\pm1}$ is dominant in this connected component
    and thus solutions on this component do not have $180^\circ$
    rotational symmetry.
    Symmetry breaking occurs via subcritical pitchfork bifurcation
    at the point ${\tt Z211}j$ to ${\tt Z212}j$ and ${\tt A212}j$ emerge.
    ${\tt A212}j$ is linked by ${\tt A211}$ via saddle-node bifurcations.
    ${\tt A211}$ is an important hub saddle and ${\tt Z210}$ (SP) is one of
    the minimizers.
  }\label{bifAtlasTeX3.eps}
\end{figure}
\ref{bifAtlasTeX3.eps}, and
\begin{figure}
  \hfil\includegraphics[width=\hsize]{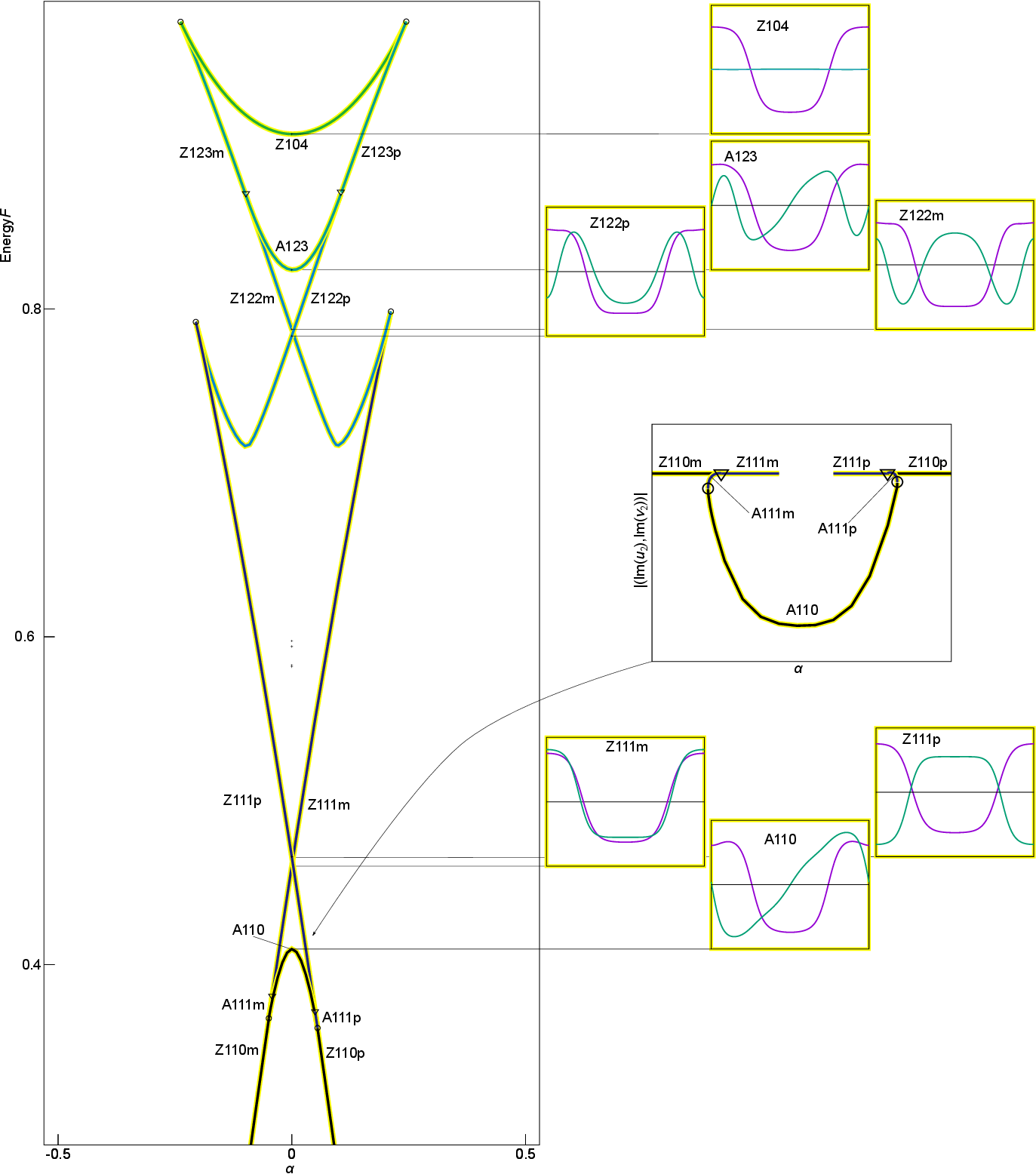}
  \caption{The fourth connected component,
    one-hump component.
    Fourier coefficients $u_{\pm1}$ is dominant in this connected component
    and thus solutions on this component do not have rotational symmetry.
    ${\tt A123}$ and ${\tt A111}j$ emerge via subcritical pitchfork
    bifurcations.
     ${\tt A110}$ (JP) is one of the minimizers.
    ${\tt Z111}j$ is a hub saddle that sorts trajectories into either the left or right Janus solution.
   }\label{bifAtlasTeX4.eps}
\end{figure}
\ref{bifAtlasTeX4.eps}, respectively.
The first connected component consists of constant solution and initial
bifurcations from that.
This is mainly understood by dispersion relation corresponding to the
constant solution and thus this component is called as
``linear regime component''.
The second connected component is called as ``two-hump component'' because
$u_2$ is dominant in all solutions in this component.
Note that, only in ${\tt D026}$, $u_2$ is small when $|\alpha|\sim0$,
however, this can be understood as $u_2$ dominant because
$u_2$ becomes dominant again in the both ends of this branch.
The third connected component is called as
``one-mode microphase component'' because
$v_1$ is dominant in all solutions in this component.
The fourth connected component is called as ``one-hump component''
by the same reason as the second component.

Whole bifurcation structure contains only 4 types of bifurcations:
Saddle-node bifurcation, super-- and sub--critical pitchfork
bifurcations, and point where Morse index changes.
These points are shown by symbols of circle,
triangle, inverse triangle, and rhombus, respectively.
See also Fig.~\ref{bifAtlasTeXLegend.eps}.

\subsubsection{Linear regime component (the first connected component)}
Magnified bifurcation diagram is shown in Fig.~\ref{bifAtlasTeX1.eps}.
This component consists of constant solution ${\tt D00}Mj$ and the branch
bifurcated from it.

There occurs a supercritical pitchfork branch at $|\alpha|\sim0.5$ and
there appears ${n}=2$-branch (two-mode) for $u$ and $v$.
This mode appears at any point because ${\tt D00}Mj\in D_\infty$ and the
Morse index $M$ changes $2$, because there always appear
$\oppi/2$-rotated branch simultaneously.
These branches do not grow and not connected to ${\tt D225}j$
in the second connected component.
The same thing occurred for ${\tt Z113}j$ and ${\tt Z111}j$.

For $\alpha\sim0.5$, a solution branch with $u\sim v$ bifurcates from the
constant solution, whereas for $\alpha\sim-0.5$, a branch with $u\sim -v$
emerges.
These branches are connected via the symmetric branch ${\tt D206}$,
along which $v\sim0$ holds at $\alpha=0$.
Near $\alpha=0$, $v$ appears with opposite sign to $u$ for $\alpha<0$,
and with the same sign for $\alpha>0$.
This connection links the branches labeled $j={\tt m}$ and $j={\tt p}$.
Although $b=0$ on ${\tt D206}$ for $\alpha\sim0$, it reaches $b=2$ when
$|\alpha|\gg0$.
Nevertheless, we focus on the behavior near $\alpha\sim0$,
and name the branch $b=0$.

\subsubsection{Two-hump component (the second connected component)}
Magnified bifurcation diagram is shown in Fig.~\ref{bifAtlasTeX2.eps}.
The solution with the largest energy in this connected component
is ${\tt D026}$.
As shown in Fig.~\ref{bifAtlasTeX2.eps}, the dominant mode
of $u$ at $\alpha=0$ is ${n}=4$, albeit small in amplitude,
so one might consider labeling it as ${\tt D426}$.
However, this ${n}=4$ mode appears only near $\alpha=0$ and is quickly
replaced by ${n}=2$ thereafter.
Furthermore, similar to ${\tt D206}$ in the linear regime component,
the solution component $u_2$ emerges around $\alpha \sim 0$ with opposite
sign depending on the sign of $\alpha$.
Therefore, we disregard the small value $a=4$ and adopt $a=0$ as the label.

The branch ${\tt D026}$ undergoes saddle-node bifurcations in both
positive and negative directions of $\alpha$, eventually becoming ${\tt D225}j$.
The phase relationship between $u$ and $v$ is the same as that in ${\tt D228}j$.
Pitchfork bifurcations occurs at the transition from ${\tt D224}j$ to
${\tt D225}j$, but this particular bifurcation is asymmetric
with respect to $\alpha=0$ in structure.
This asymmetry is attributed to the nonzero mean values $\bar u,\bar v\neq0$.

From the point where ${\tt D224m}$ transitions to ${\tt D225m}$,
a new branch ${\tt C225m}$ emerges via a subcritical pitchfork bifurcation.
Likewise, from the transition point of ${\tt D224p}$ to ${\tt D225p}$,
the branch ${\tt C224p}$ appears via a supercritical pitchfork bifurcation.
In this symmetry breaking, the reflection symmetry is lost,
while the $180^\circ$ rotational symmetry is preserved.
The branch ${\tt C225m}$ immediately becomes ${\tt C224m}$,
and from this point, ${\tt A224m}$ appears via a supercritical pitchfork
bifurcation.
On the other hand, ${\tt C224p}$ immediately transitions to ${\tt C223}$,
from which ${\tt A224p}$ appears via a subcritical pitchfork bifurcation.
The branch ${\tt C223}$ connects to ${\tt C224m}$ through a saddle-node
bifurcation.
When the bifurcation proceeds from $C$ to $A$, even the remaining
$180^\circ$ rotational symmetry is lost.

Next, when ${\tt D222}j$ transitions to ${\tt D223}j$, a new branch
${\tt Z213}j$ emerges via a subcritical pitchfork bifurcation.
In contrast to the previous symmetry breaking, this time the
$180^\circ$ rotational symmetry is lost, while the reflection symmetry is
preserved.
The branch ${\tt Z213}j$ becomes ${\tt Z214}j$ and the two are connected by
two saddle-node bifurcations involving ${\tt Z213}$.
Just after the bifurcations from ${\tt D222}j$ and ${\tt D223}j$,
${\tt Z213}j$ is dominated by $b = 2$, but
${n}=1$ mode begins to grow from the bifurcation point.
As a result, the $180^\circ$ rotational symmetry is broken,
and as the solution branch moves away from the bifurcation point,
the ${n}=1$ mode gradually replaces ${n}=2$ as the dominant one.
Therefore, we adopt $b = 1$ as the label for this branch.

${\tt D222}j$ continues to larger $|\alpha|$ at least $|\alpha|\le0.53$.

\subsubsection{One-mode microphase component (the third connected component)}
Magnified bifurcation diagram is shown in Fig.~\ref{bifAtlasTeX3.eps}.
The solution with the highest energy in this connected component is
${\tt Z014}$.
Although $a=2$ at $\alpha=0$, the solution undergoes a transition in which
a solution with $a=1$ becomes dominant after passing through
a saddle-node bifurcation.
However, since $|u_2|$ remains significantly smaller than $|v_1|$,
we regard this state as $a=0$ in the present context.
The branch ${\tt Z014}$ lacks $180^\circ$ rotational symmetry and possesses
only reflection symmetry.
The branch ${\tt Z014}$ connects via saddle-node bifurcations at both ends
to ${\tt Z113}j$, and immediately $u_2$ emerges, leading to ${\tt Z212}j$.

At the transition point from ${\tt Z212}j$ to ${\tt Z211}j$, a subcritical
pitchfork bifurcation gives rise to ${\tt A212}j$, where the reflection
symmetry is lost.
The branch ${\tt Z211}j$ connects to ${\tt Z210}$, and
the branch ${\tt A212}j$ connects to ${\tt A211}$, respectively
via saddle-node bifurcations.

The SP solution ${\tt Z210}$ is a stable state near $\alpha=0$ and
represents one of the final outcomes of the solution transition.

\subsubsection{One-hump component (the fourth connected component)}
Magnified bifurcation diagram is shown in Fig.~\ref{bifAtlasTeX4.eps}.
The solution with the highest energy in the fourth connected component is
${\tt Z104}$.
This branch connects to ${\tt Z123}j$ via saddle-node bifurcations at both ends.
As in other components, the sign of $v$ depends on the sign
of $\alpha$.
This branch proceeds through ${\tt Z122}j$ and reaches ${\tt Z111}j$.
At the bifurcation point from ${\tt Z123}j$ to ${\tt Z122}j$, a subcritical
pitchfork bifurcation occurs, breaking reflection symmetry and giving rise
to ${\tt A123}$.
This branch connects ${\tt m}$ and ${\tt p}$.

Up to ${\tt Z122}j$, the $b=2$ mode is dominant.
However, as the branch passes through a saddle-node bifurcation and
transitions to ${\tt Z111}j$, the $b=1$ mode gradually becomes dominant.
By the time it further bifurcates into ${\tt Z110p}$ and ${\tt Z110m}$,
the $b=1$ mode fully dominates.

From this bifurcation point, a subcritical pitchfork bifurcation gives
rise to ${\tt A111}j$, at which the reflection symmetry is lost.
${\tt A111}j$ connects via a saddle-node bifurcation to ${\tt A110}$.

The solution ${\tt A110}$ is another stable state near $\alpha=0$,
and represents an important alternative final state.
In the narrow range $0.04988\le\alpha\le0.05523$, both ${\tt A110}$ and
${\tt Z110p}$ are stable.
Likewise, in the range $-0.049097\le\alpha\le-0.04186$,
both ${\tt A110}$ and ${\tt Z110m}$ are stable.
Together with the region $-0.0207\le\alpha\le0.024$, where both
${\tt Z210}$ and ${\tt A110}$ are stable,
they form bistable regions.

\subsection{Saddle network}\label{Sec:Saddlenetwork}
The approximated unstable manifold is obtained calculating the
time evolution from initial values which are made adding small
eigenfunctions $(\hat u,\hat v)$ to the solution $(u,v)$ with $r = 1$.
The amplitude of $(\hat u,\hat v)$ is controlled as
$|(\hat u,\hat v)|=0.01$.
Note that $O(|(u,v)|)=1$.

Information of network among saddles gives a skeleton of transition process
from constant solution ${\tt D00}Mj$ to the minimizers $Xab{\tt 0}j$.
As is mentioned above, global bifurcation diagram gives an atlas of the
landscape.
In the same sense, saddle network gives a ``road map'' among towns
($=$saddles).
Particularly, connection among ``hub saddles'' give an ``arterial road map''
of the land.
The hub saddle is a saddle which has codimension 1 in an appropriate invariant subspace.
Therefore stable manifold comprises boundary of
asymptotic states in the subspace and one dimensional unstable manifold
plays the role of ``hub''.
In other words, if $(u,v)$ is a hub saddle, there exists an eigenfunction
$(\hat u,\hat v)$ s.t.
$(u,v)+\varepsilon(\hat u,\hat v)$ and
$(u,v)-\varepsilon(\hat u,\hat v)$ belong to different basin
in the subspace when $\varepsilon\to+0$.

List of hub saddles are given in Table
\begin{table}[htbp]
  \centering
  \begin{tabular}{lrccr}
    \hline
    Saddle
    &\begin{minipage}{10ex}Eigenvalue\end{minipage}&\begin{minipage}{12ex}Symmetry of subspace\end{minipage}&\begin{minipage}{12ex}Symmetry of eigenfunction\end{minipage}&\begin{minipage}{12ex}Asymptotic states $+\varepsilon$/$-\varepsilon$\end{minipage}\\\hline\hline
    $\tt{Z213 }$&1.492&Z&Z&${\tt Z111m}$/${\tt Z111p}$\\\hline
    $\tt{Z212m}$&4.748&Z&Z&${\tt Z111p}$/${\tt Z210 }$\\\hline
    $\tt{Z212p}$&2.440&Z&Z&${\tt Z210 }$/${\tt Z111m}$\\\hline
    $\tt{Z111p}$&0.634&A&A&${\tt A110 }$/${\tt A110 }$\\\hline
    $\tt{Z111m}$&0.824&A&A&${\tt A110 }$/${\tt A110 }$\\\hline
    $\tt{A211 }$&0.334&A&A&${\tt Z210 }$/${\tt A110 }$\\\hline
  \end{tabular}
  \caption{
    List of hub saddles corresponding to Fig.~\ref{fig:hubsaddle.eps}.
    Although asymptotic state from ${\tt Z111}j$ is the same minimizer ${\tt A110}$, symmetry of ${\tt A110}$ is different as shown in Fig.~\ref{fig:hubsaddle.eps}.
    Although the Morse indices of ${\tt Z213 }$ and ${\tt Z212}j$ are not 1, those in the subspace of ${\tt Z}$-symmetry are 1.
  }
  \label{tb:hub}
\end{table}
\ref{tb:hub} and connections among them are given in Fig.
\begin{figure}
  \hfil\includegraphics[width=.9\hsize]{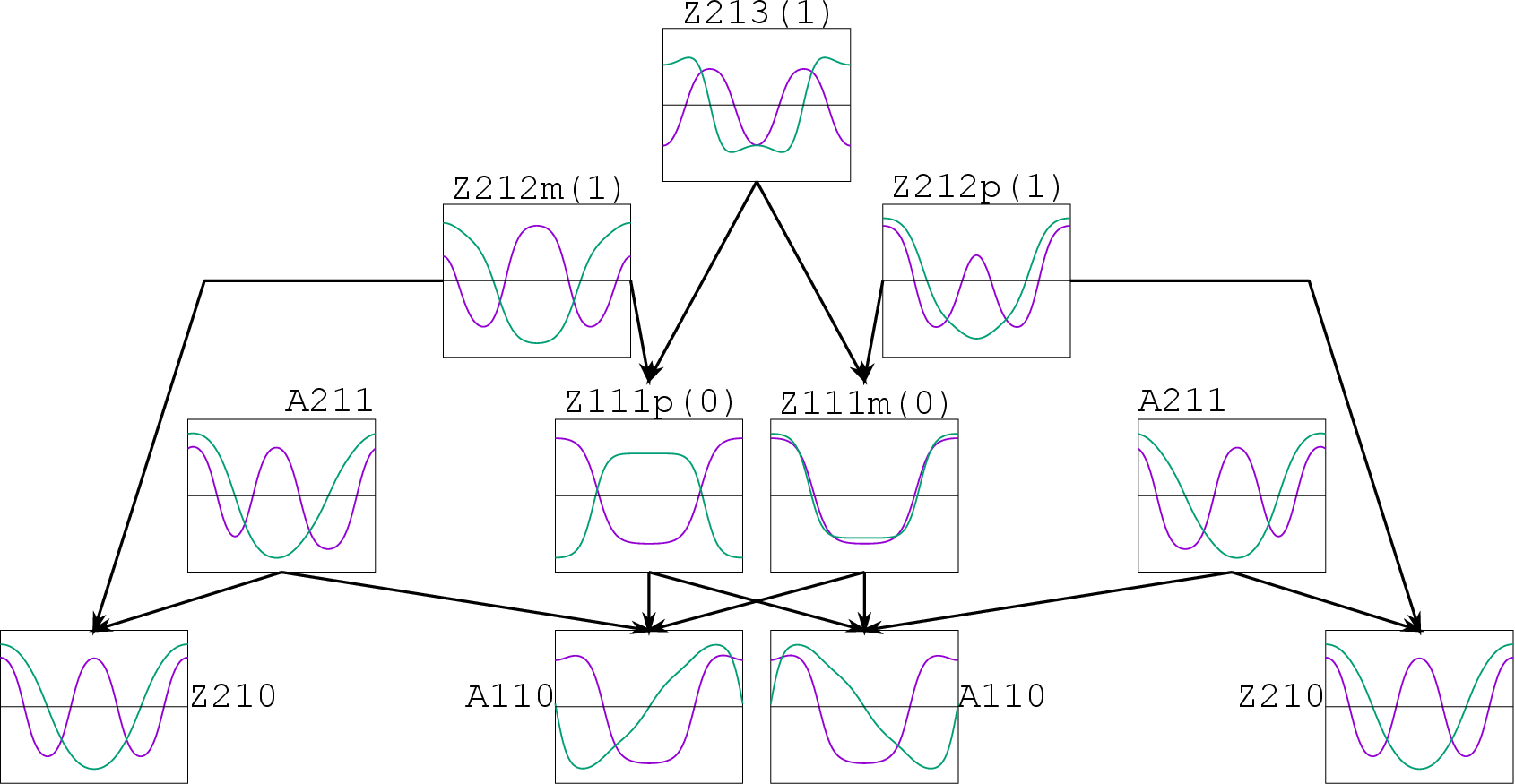}
  \caption{
    The connections among hub saddles and minimizers corresponding to Table \ref{tb:hub}.
    There exist six hub saddles and two minimizers.
    Codimension in the ${\tt Z}$-symmetric subspace is shown in parentheses.
    Solutions are stratified in ascending order of Morse index and thus
    transitions occur from top to bottom.
  }
  \label{fig:hubsaddle.eps}
\end{figure}
\ref{fig:hubsaddle.eps}.

The specific transition paths with respect to different initial values
are discussed in the next section.
In these discussions, the relationship between transition paths and
three maps --- atlas, road map, and arterial road map --- is given
in detail.

In the rest of this section, properties of two important saddles for
the case $\alpha=0.01$ are considered.

\subsubsection{${\tt D008}$}\label{section:D008}
Codimension of the unstable manifold of the constant solution ${\tt D008}$
is $8$.
However, because ${\tt D008}$ has the symmetry $D_\infty$,
if a function is an eigenfunction, then any arbitrary translation of that
function is also an eigenfunction.
Because any eigenfunction always has this degree of freedom of
translation, it always appears with a phase-shifted pair and thus,
intrinsic codimension is 4.
The dominant mode of these four eigenfunctions are
$(\hat u_2,\hat v_2)$, $(\hat u_2,-\hat v_2)$, $(\hat u_1,\hat v_1)$, and
$(\hat u_1,-\hat v_1)$,
which correspond to
${\tt D225p}$, ${\tt D225m}$, ${\tt Z111p}$, and ${\tt Z111m}$, respectively.
The eigenvalues of them are
$12.646$, $12.323$, $5.339$, and $5.258$, respectively.
The first, second, third and fourth eigenvalues are slightly
different due to the non-zero values of $\overline{u}$ and $\overline{v}$.
Although the first and the second eigenvalues are different, they are
almost identical because $\overline{u}$ and $\overline{v}$ are small
and thus superposition of the translation of
first and second eigenfunctions becomes dominant.
Typical directions emerged from this superposition is the following
five directions:
\begin{itemize}
\item
  $\hat u_2\sim\hat v_2$.
\item
  $\hat u_2\sim-\hat v_2$.
\item
  $\hat u_2$ is dominant and $\hat v_2=0$.
\item
  $\hat v_2$ is dominant and $\hat u_2=0$.
\item
  $|\hat u_2|\sim|\hat v_2|$ and the phase difference between them
  is $\oppi/2$.
\end{itemize}
Each of the above direction corresponds to
${\tt D225p}$, ${\tt D225m}$, ${\tt D206}$, ${\tt D026}$, and ${\tt C223}$,
respectively and therefore all emerged paths for $r=1$, which is
shown in the subsequent section, is explained as the unstable
manifold of ${\tt D008}$.

\subsubsection{${\tt C223}$}\label{section:C223}
As is shown in the next section, ${\tt A211}$ is a hub saddle.
Its stable manifold is the basin boundary of ${\tt A110}$ and ${\tt Z210}$.
Here the ratio of the relaxation parameters $r$ plays an
essential role. The $r$-dependence of the asymptotic state is investigated in the
next section comprehensively.

When $r=1$, ${\tt C223}$ exists at the side of ${\tt A110}$ and
thus three dimensional unstable direction of ${\tt C223}$ is connected
to the stable manifold of ${\tt A110}$.
The solutions themselves do not change, however the stable manifold
of ${\tt A110}$ changes and at the point $r=r^*\sim1.41$, heteroclinic
connection between ${\tt C223}$ and ${\tt A211}$ emerges.
After this reconnection, ${\tt C223}$ goes to the other side, the
side of ${\tt Z210}$, and thus three dimensional unstable direction of
${\tt C223}$ is connected
to the stable manifold of ${\tt Z210}$.
This process of reconnection is shown in
\begin{figure}
  \hfil\includegraphics[width=.9\hsize]{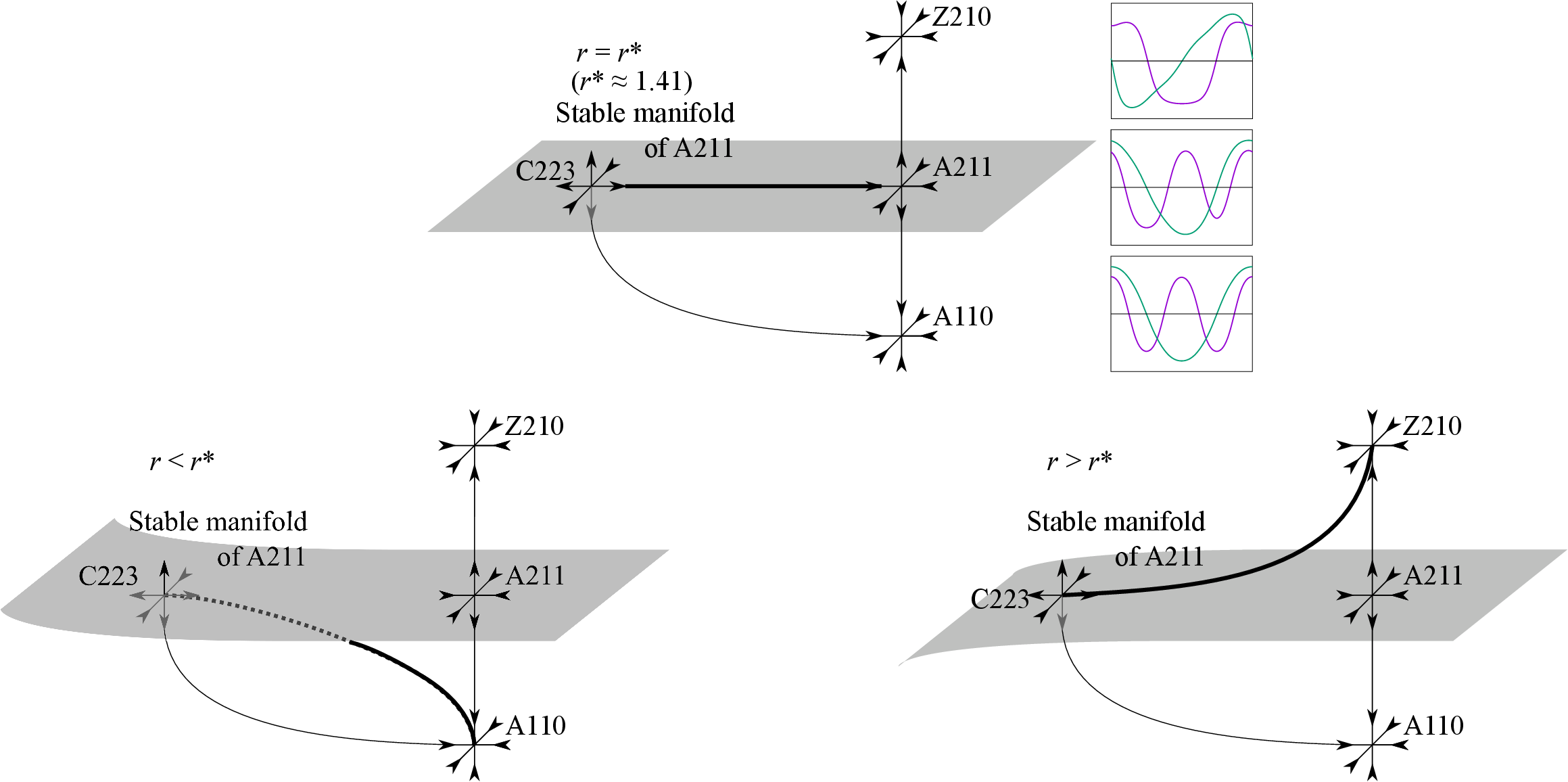}
  \caption{
    $r$-dependence of the stable manifold of ${\tt A211}$.
    ${\tt C223}$ is an important saddle because most of flows from
    ${\tt D008}$ with small random perturbations pass close to that
    when $r\sim1$.
    After passing ${\tt C223}$, many flows are sorted by ${\tt A211}$
    and go either ${\tt A110}$ or  ${\tt Z210}$.
    Therefore the stable manifold of ${\tt A211}$ plays the essential
    role.
    As shown in Table \ref{tb:statistics_r1}, almost all flows
    approache ${\tt A110}$ when $r=1$.
    On the other hand, as shown in Table \ref{tb:statistics_r1.38},
    a half of the flows approache ${\tt A110}$ whereas the other half
    approache ${\tt Z210}$ when $r=1.38$.
    The difference between them is attributed to the relationship
    between the stable manifold of ${\tt A211}$ and the saddle ${\tt C223}$.
  }
  \label{fig:C223-A211.eps}
\end{figure}
Fig.~\ref{fig:C223-A211.eps}.
This reconnection process accounts for the change of paths and
asymptotic states depending on the parameter $r$ which is
shown in detail in the next section. Recall that the profile of the free energy remains the same as $r$ varies, but the trajectories are changed as described above.

\section{Classification of trajectory behaviors}

Describing the evolution of trajectories on a rugged free energy landscape with multiple metastable states is inherently challenging. This difficulty stems from the simultaneous influence of numerous local minima and an even greater number of saddle points distributed along the connecting pathways. Consequently, even when initial condition is fixed, small random perturbations can drastically affect the trajectory's path. In such settings, guiding the system toward a desired morphology is highly nontrivial and often relies on trial and error.
Our aim is to mitigate this uncertainty and, building on the saddle network analysis presented earlier, to clarify how the intermediate dynamics can be systematically steered by adjusting the relaxation parameter ratio $r$. To this end, we take as initial data a perturbed version of the constant solution $\tt{D008}$, generated by superimposing small random fluctuations, and investigate the dynamics of the solution under the numerical scheme and parameter settings introduced in Sec. 2.
The random perturbations are uniformly distributed between $0$ and $0.001$ for $u$ and between $0.005$ and $0.006$ for $v$ throughout this section.
Notably, the effect of random perturbations becomes negligible when $r \ll 1$ or $r \gg 1$, enabling robust control over the entire dynamical evolution.

\subsection{Transition dynamics and their dependence on the parameters}
\begin{figure}[htbp]
  \begin{minipage}[b]{0.33\linewidth}
    \centering
    \includegraphics[keepaspectratio, scale=0.8]{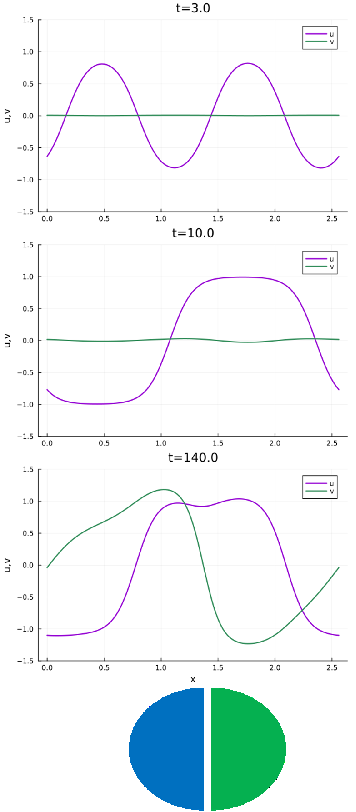}
    \subcaption{$r=0.01$}
  \end{minipage}
  \begin{minipage}[b]{0.33\linewidth}
    \centering
    \includegraphics[keepaspectratio, scale=0.8]{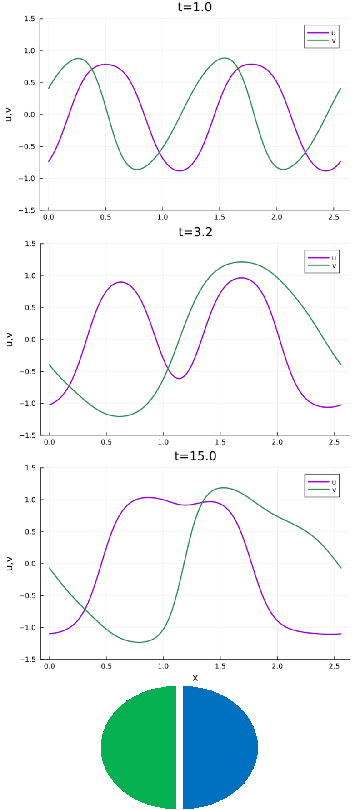}
    \subcaption{$r=1$}
  \end{minipage}
  \begin{minipage}[b]{0.33\linewidth}
    \centering
    \includegraphics[keepaspectratio, scale=0.8]{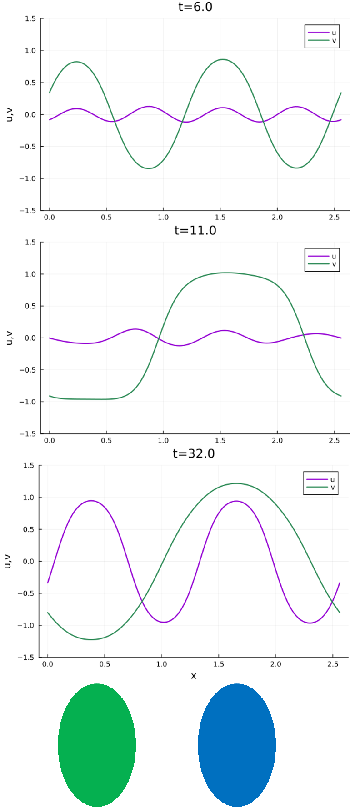}
    \subcaption{$r=100$}
  \end{minipage}
  \caption{
    Examples of the dynamics of the CCH system $(u,v)$ with $r=0.01, 1, 100$ along with conceptual diagrams of their final steady states. Purple lines show $u=u(x)$, and green lines show $v=v(x)$. Time advances from top to bottom, with bottom row representing the final steady state. For $r=0.01$ and $r=1$, the solutions converge to the same profile, $\tt{A110}$ (JS). They are mirror images of each other. For $r=100$, the solution converges to a different minimizer, $\tt{Z210}$ (SP). When $r = 0.01 < 1$, $u$ evolves faster than $v$; when $r = 100 > 1$, $v$ evolves faster than $u$.
  }
  \label{fig:dynamics}
\end{figure}

We already see in the previous section that the ratio $r := \tau_u/\tau_v \in (0,\infty)$ affects the dynamics of trajectories. The smaller the ratio $r$, the faster $u$ dynamics; the larger the ratio, the faster $v$ dynamics. For example, Fig.~\ref{fig:dynamics} shows the dynamics of solutions for $\alpha = 0.01$ with random perturbations added to the initial constant solution $\tt{D008}$. At $r=0.01$, $u$ relaxes first forming one--hump solution by $t=10.0$ followed by 
$v$ evolving into the JP solution at $t=140.0$ (see panel (a)). At $r=100$, the development order of $u$ and $v$ is reversed, yielding the SP solution (see panel (c)).
On the other hand, when $r=1$, predicting the final outcome is more intricate because both 
$u$ and $v$ deform simultaneously. At $t=15.0$, it is still unclear whether the trajectory will relax to the JP solution or the SP solution. Although the profile at $t=15.0$ resembles the SP solution, it ultimately settles into the JP solution. Here one of the hub saddles $\tt{A211}$ plays a key role at the last stage of evolution, and the final outcome depends on a random seed.

Inspired by this observation, it is convenient to visualize the locus of trajectory and locate saddles in the following space $(F_1, F_2)$ consisting of energy contributions from $u$ and $v$, respectively.
Here, $(F_1,F_2)$ space is defined by the decomposition $F[u,v] = F_1[u,v] + F_2[u,v]$ as follows:
\begin{gather}
  F_1[u,v] := \int_0^L \left\{  \frac{\epsilon_u^2}{2}\left| \frac{\partial u}{\partial x} \right|^2 + \frac{(1-u^2)^2}{4} + \frac{\alpha}{2}uv + \frac{\beta}{2}uv^2 \right\}\dif x,
  \\
  F_2[u,v] := \int_0^L \left\{ \frac{\epsilon_v^2}{2}\left| \frac{\partial v}{\partial x} \right|^2 + \frac{(1-v^2)^2}{4} + \frac{\alpha}{2}uv + \frac{\beta}{2}uv^2 \right\}\dif x.
\end{gather}

\subsubsection{How the energy landscape depends on $\alpha$}
\begin{figure}[htbp]
  \centering
  \begin{tabular}{cc}
    \multicolumn{2}{c}{
      \begin{subfigure}[t]{\linewidth}
        \centering
        \includegraphics[width=\linewidth]{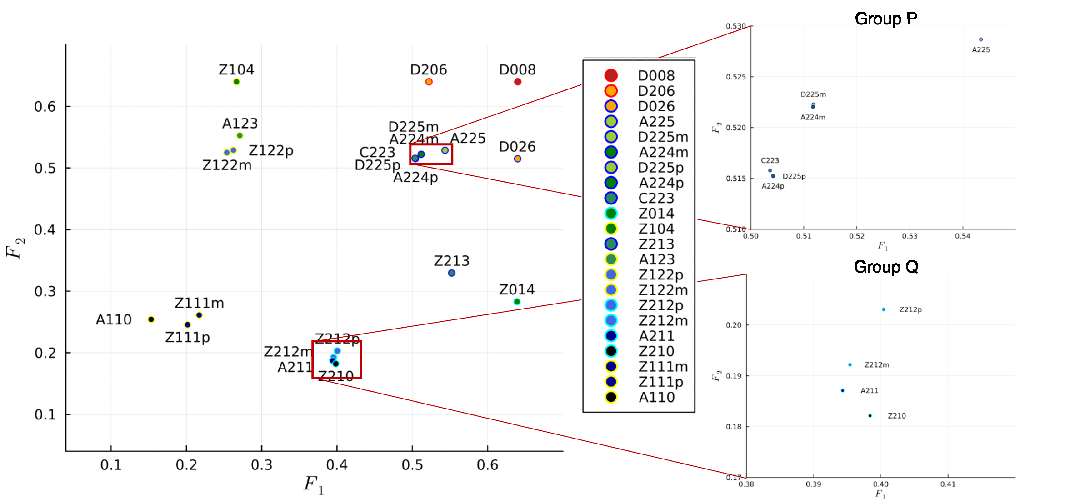}
        \subcaption{$\alpha=0.01$}
        \label{fig:energyLandscape_a001}
      \end{subfigure}
    } 
    \\[6ex]
    \begin{subfigure}[t]{0.45\linewidth}
      \centering
      \includegraphics[width=\linewidth]{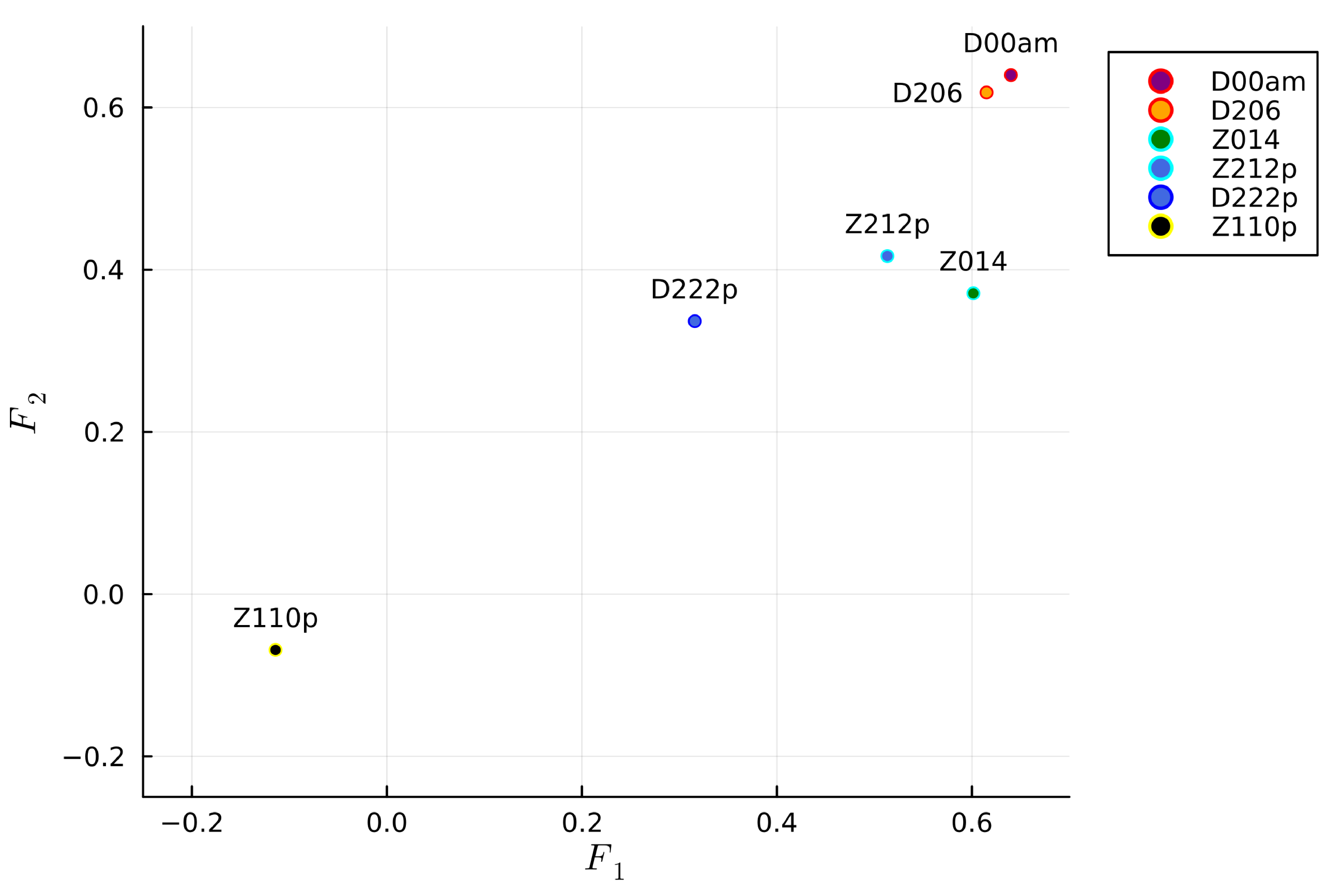}
      \subcaption{$\alpha=0.3$}
      \label{fig:energyLandscape_a03}
    \end{subfigure}
    &
    \begin{subfigure}[t]{0.45\linewidth}
      \centering
      \includegraphics[width=\linewidth]{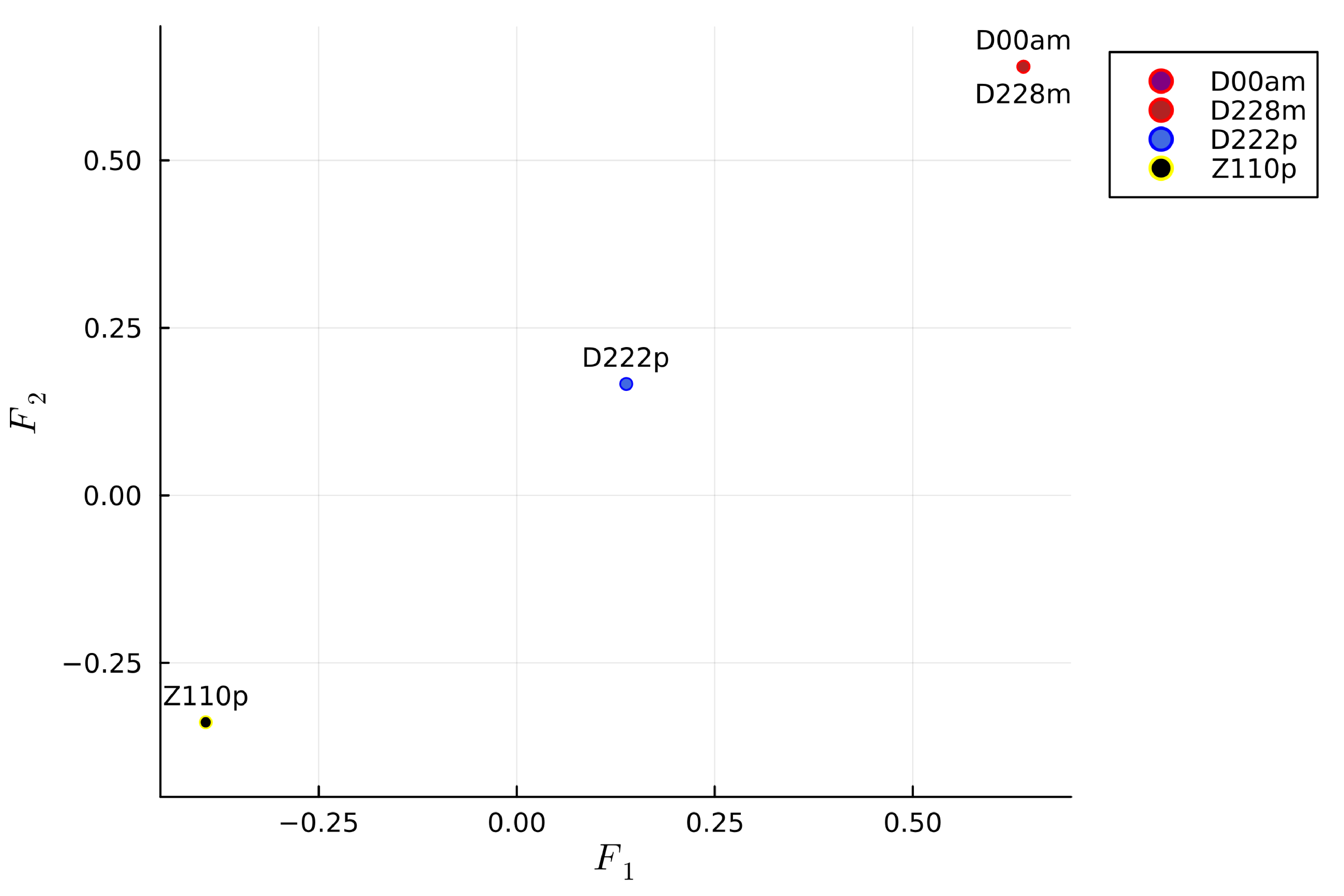}
      \subcaption{$\alpha=0.5$}
      \label{fig:energyLandscape_a05}
    \end{subfigure}
  \end{tabular}
  \caption{
    Distribution of stationary solutions in energy space $(F_1,F_2)$.
    Three cross-sections of Fig.~\ref{bifAtlasTeX.eps} at $\alpha = 0.01$, $0.3$, and $0.5$ are shown.
    The energy landscape for $\alpha \approx 0$ represents the most complex case. As the magnitude of $|\alpha|$ increases, the number of stationary solutions (plots) decreases, and the energy landscape becomes simpler. The inlets consists of two saddle clusters, P and Q. }
  \label{fig:energyLandscape}
\end{figure}

As shown in the global bifurcation diagram in Fig.~\ref{bifAtlasTeX.eps}, the energy landscape exhibits substantial changes with variations in the parameter $\alpha$.
In particular, when $|\alpha|$ is large, the total number of saddle points decreases significantly, and the only remaining local minimizer is ${\tt Z110}j$, resulting in a simplified energy landscape.
For example, Fig.~\ref{fig:energyLandscape} shows the distribution of stationary solutions in the $(F_1, F_2)$  space for $\alpha = 0.01, 0.3, 0.5$. As is expected, the number of solutions is decreased as $\alpha$ is increased, which suggests much less freedom for trajectory behaviors. On the other hand, at $\alpha=0.01$ (Fig.~\ref{fig:energyLandscape_a001}), the enlarged view shows a dense cluster of saddles with dominant wave number $2$ for both $u$ and $v$ (Group P), stationary solutions with dominant wave number $2$ for $u$ and $1$ for $v$ (Group Q). These clusters attract most trajectories and channel them toward the next stage, as described in the following section.

\begin{figure}[htbp]
  \hfil\includegraphics[width=0.9\linewidth]{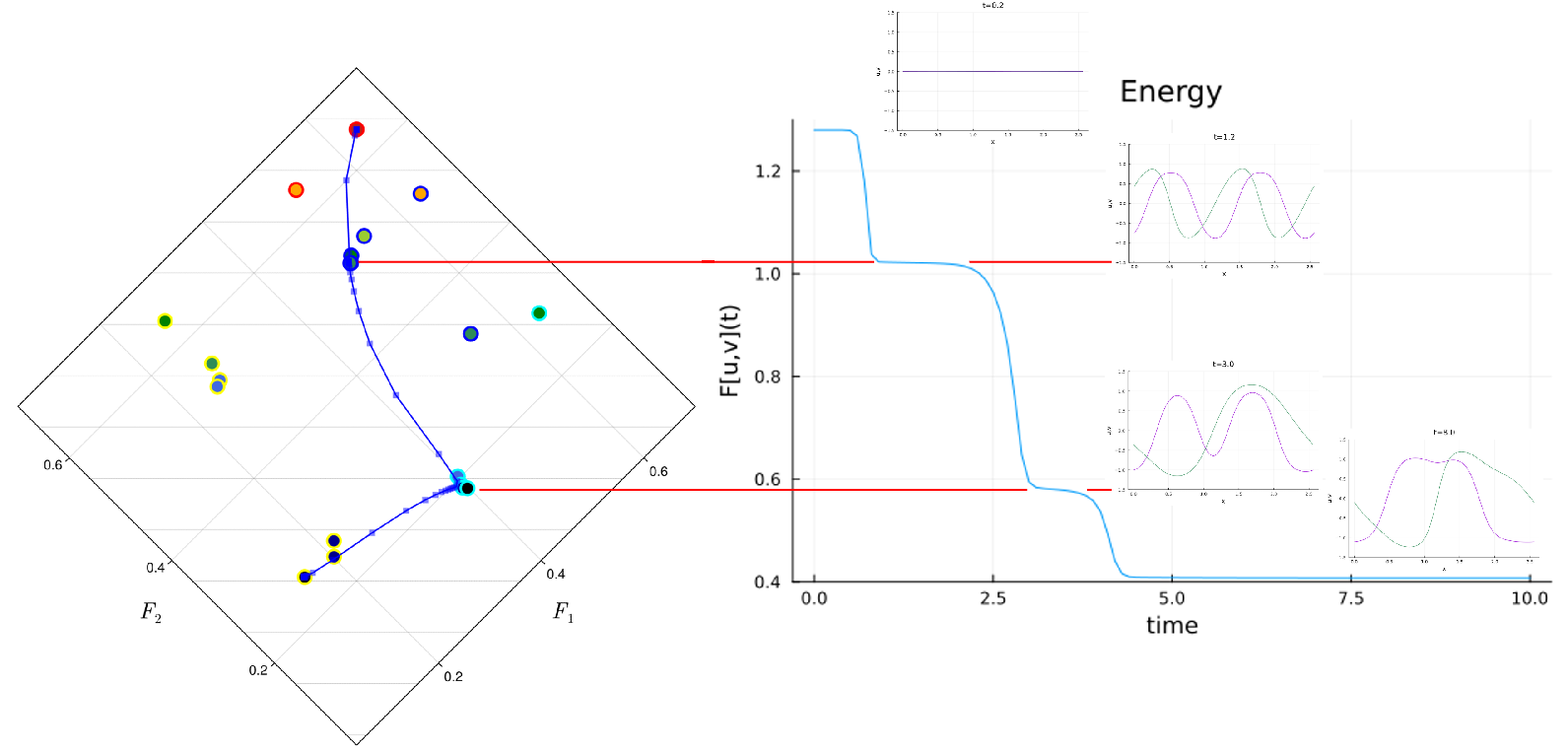}
  \caption{
    A numerical solution for a given initial datum with $\alpha=0.01$ and $r=1$:
    The trajectory in energy space $(F_1,F_2)$ (left; blue polyline) and time evolution of energy (right) are shown. The solution profiles of $(u,v)$ at energy plateau are close to the saddle points in Fig.~\ref{bifAtlasTeX.eps}. The left panel is rotated by $45^\circ$.
  }
  \label{fig:dynamics_energy_flow}
\end{figure}

A typical trajectory for $\alpha=0.01$ and $r=1$ is shown in Fig.~\ref{fig:dynamics_energy_flow}.
The left side of Fig.~\ref{fig:dynamics_energy_flow} shows the flow of energy of the solution in the $(F_1, F_2)$ energy space, while the right side shows the decay of the total energy $F=F_1+F_2$.
In addition, the solution $(u,v)$ at each plateau is indicated, which is close to the saddle nearby.

\subsubsection{Expansion and contraction of trajectory bundles}
Now we are ready to consider how the trajectories behave starting from an initial data $\tt{D008}$ with random fluctuation. We prepare 20 different random seeds for fluctuation and keep tracking the trajectories. We first discuss how the ``bundle'' of these trajectories expands or contracts depending on the parameters, rather than focusing on each trajectory.
Next, we discuss how the saddle network introduced in the previous section contributes to the expansion and contraction of solution bundle in the next section.
We adopt the ratio $r$ of relaxation parameters as control parameter for a fixed $\alpha$.
By appropriately controlling the parameters, we can significantly reduce the ``expansion'' caused by the random seeds, and in fact, in a certain parameter regime, the bundle almost converges into a single trajectory.

\begin{figure}
  \hfil\includegraphics[scale=.6]{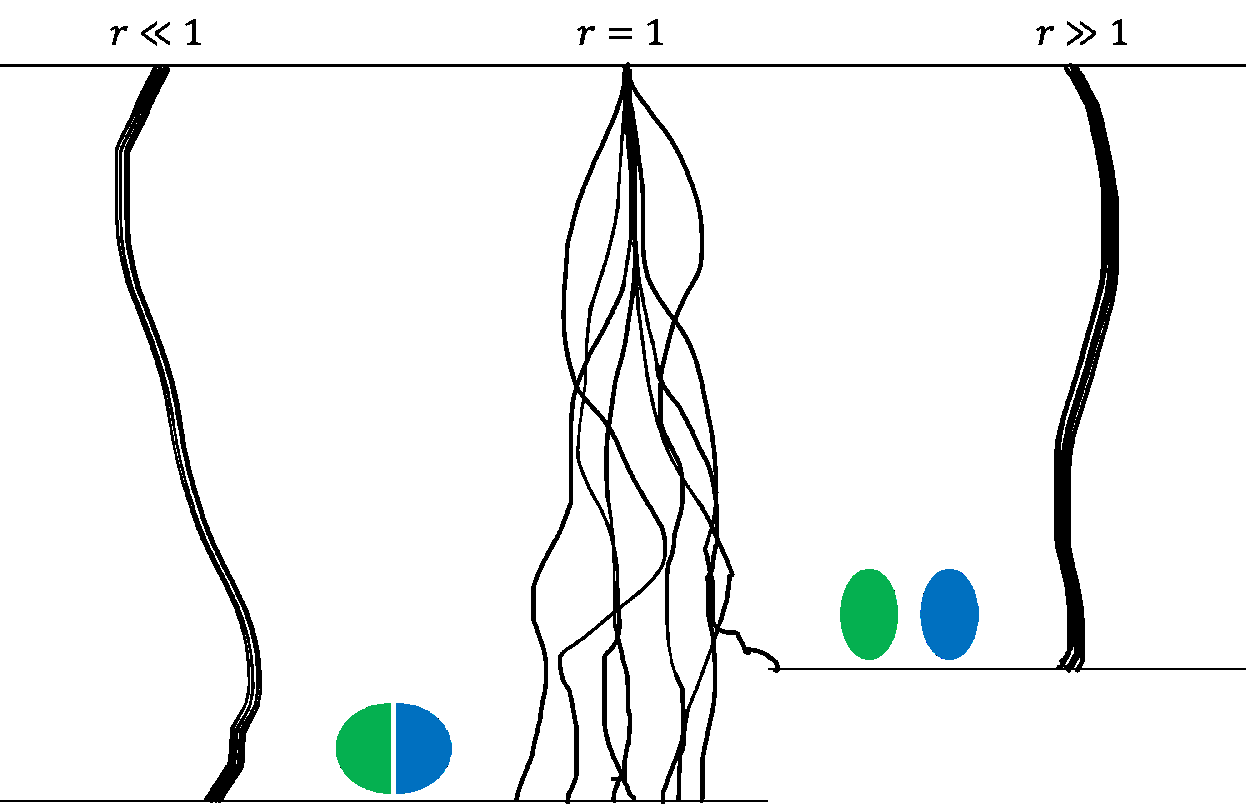}
  \caption{
    Conceptual diagram of flow bundle for three different ratios of relaxation parameters.
    When $r$ is sufficiently large or small, the trajectory bundle contract and converge to the same local minimizer.
    On the other hand, for $r \sim 1$, the bundle expands and each trajectory is sensitive to the initial random seed.
  }
  \label{fig:flow_bundle}
\end{figure}

Recall that the ratio $r$ does not alter the shape of the free energy, but instead determines how the free energy decreases along the trajectory. By tuning $r$, we can significantly reduce the expansion described above.
Specifically, the saddle points passed through are uniquely determined when $r \ll 1$ and $r \gg 1$.
Here, ``passed through'' means that the orbit passes near the saddle points without asymptotically approaching them along their stable manifolds.
If we schematically depict how the bundle of trajectories expands or contracts depending on $r$, it looks like Fig.~\ref{fig:flow_bundle}.

\begin{figure}[htbp]
  \begin{minipage}[b]{0.3\linewidth}
    \centering
    \includegraphics[keepaspectratio, width=1\linewidth]{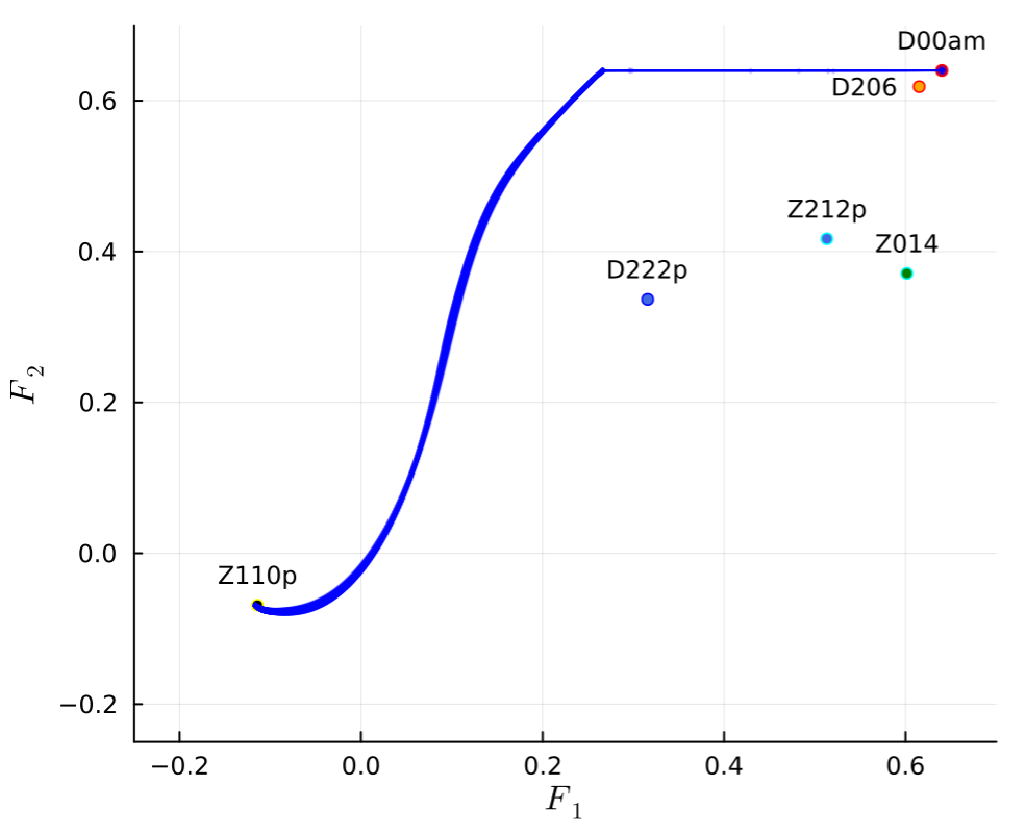}
    \subcaption{$r=0.0001$}
  \end{minipage}
  \begin{minipage}[b]{0.3\linewidth}
    \centering
    \includegraphics[keepaspectratio, width=1\linewidth]{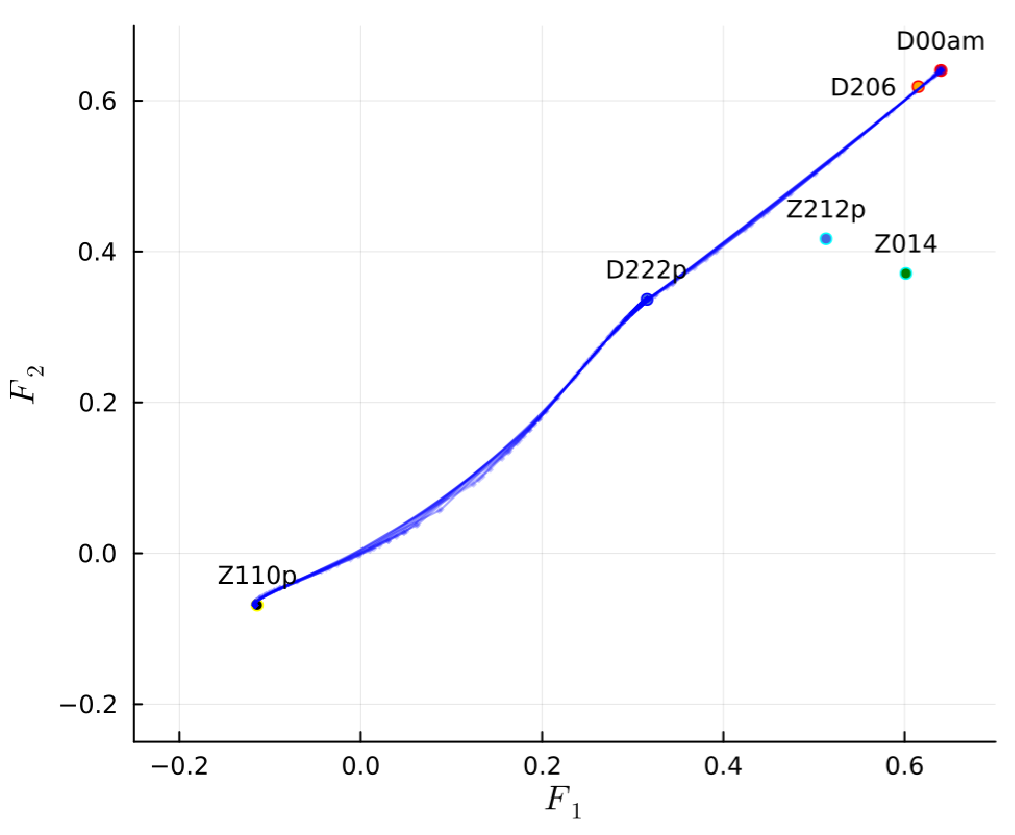}
    \subcaption{$r=1$}
  \end{minipage}
  \begin{minipage}[b]{0.35\linewidth}
    \centering
    \includegraphics[keepaspectratio, width=1\linewidth]{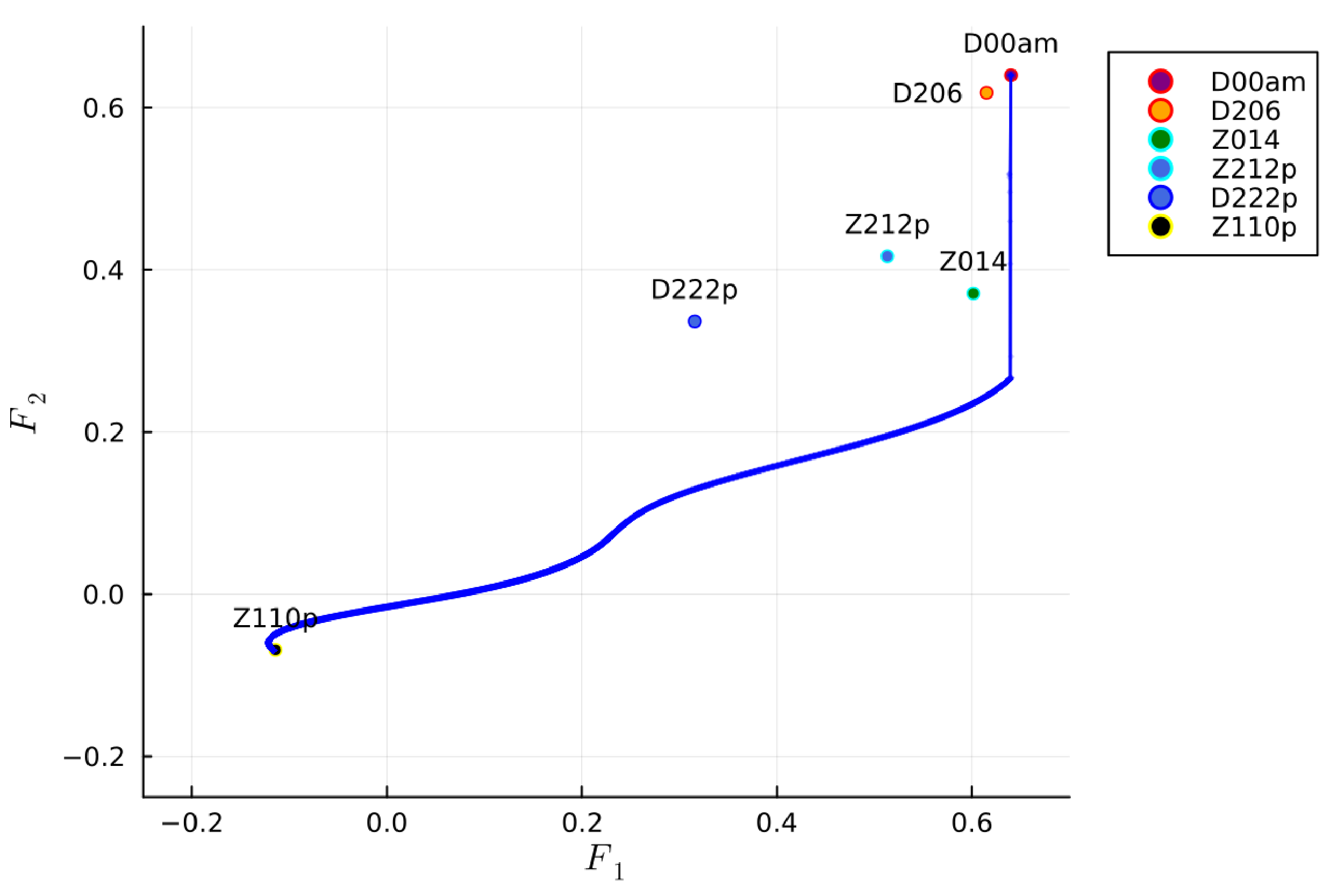}
    \subcaption{$r=10000$}
  \end{minipage}
  \caption{
    The landscape for $\alpha=0.3$ becomes much simpler than that of $\alpha \sim 0$. Trajectories for 20 initial data in the energy landscape for the ratio $r=0.0001$, $1$, $10000$ and $\alpha=0.3$.
    The flow bundles do not expand for $\alpha = 0.3$.
  }
  \label{fig:large_alpha}
\end{figure}

\paragraph{Behavior of flow bundle for large $|\alpha|$}
First, we investigate how the trajectory bundle changes with $r$ in the range where $|\alpha|$ is large. Fig.~\ref{fig:large_alpha} shows the flow of solutions in the $(F_1, F_2)$ energy space for three settings of $r=1, 0.0001, 10000$ with $\alpha=0.3$ and 20 random perturbations added to the constant solutions $\tt{D00am}$ as the initial data. The trajectory bundle does not expand and remains almost one single orbit independent of random seeds. Uncertainty arising from random seeds is reduced and well controlled.
Additionally, the trajectories spend a prolonged time at the knee points when $r=0.0001, 10000$.
The interesting thing is that any trajectory can traverse all saddle points as $r$ varies from small to large.

\begin{figure}[htbp]
  \centering
  \begin{tabular}{cc}
    \multicolumn{2}{c}{
      \begin{subfigure}[t]{\linewidth}
        \centering
        \includegraphics[keepaspectratio, scale=0.8]{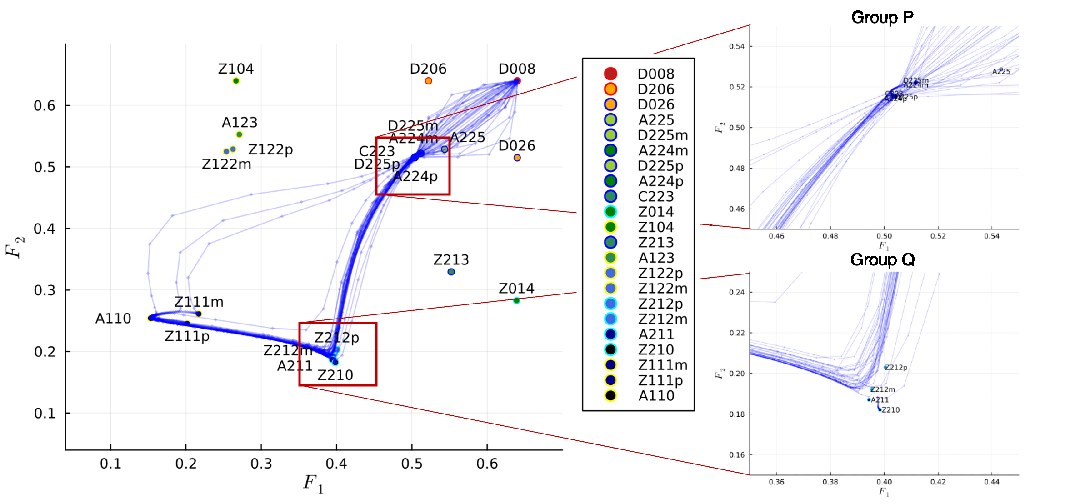}
        \subcaption{$r=1$}
        \label{fig:flow_r1}
      \end{subfigure}
    } 
    \\[1ex]
    \begin{subfigure}[t]{0.45\linewidth}
      \centering
      \includegraphics[keepaspectratio, scale=0.8]{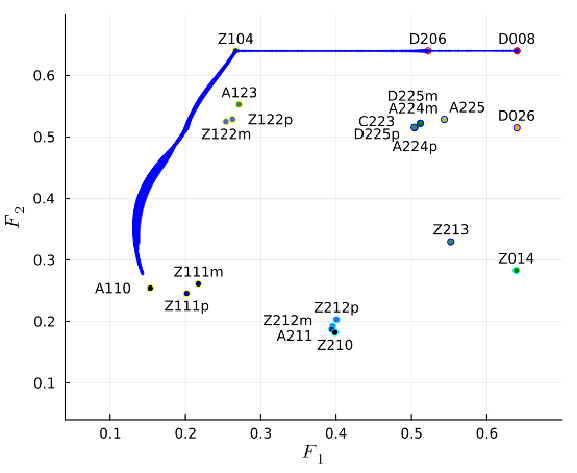}
      \subcaption{$r=0.01$}
      \label{fig:flow_r001}
    \end{subfigure}
    &
    \begin{subfigure}[t]{0.45\linewidth}
      \centering
      \includegraphics[keepaspectratio, scale=0.8]{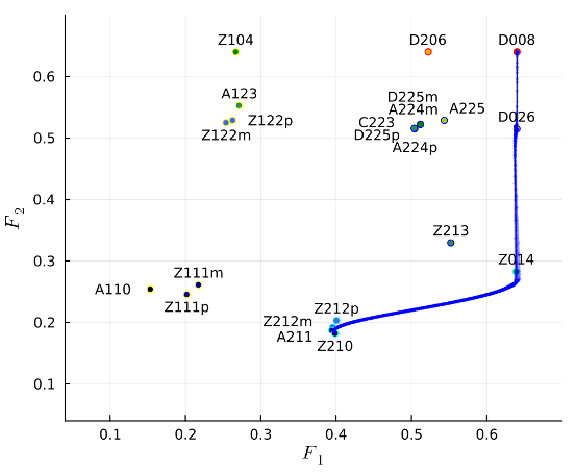}
      \subcaption{$r=100$}
      \label{fig:flow_r100}
    \end{subfigure}
  \end{tabular}
  \caption{
    The energy landscape becomes complex for $\alpha=0.01$. Trajectories are plotted for 20 initial data when the ratio $r=1, 0.01, 100$.
    The trajectory bundles do not expand for small and large $r$, however, when  $r=1$, each trajectory is very sensitive to the initial random seed, and the bundle expands at the early stage.
    Then, the trajectories are attracted to the saddle Group P in Fig.~\ref{fig:energyLandscape_a001}.
    After leaving the area P, most trajectories go to saddle Group Q area. An examination of Group Q reveals that one orbit converges to $\tt{Z210}$, whereas the others deviate from it and head toward $\tt{A110}$.
  }
  \label{fig:small_alpha}
\end{figure}

\paragraph{Behavior of flow bundle for small $|\alpha|$}
The energy landscape in $(F_1, F_2)$ space becomes complex when $\alpha \sim 0$ (here, we set $\alpha=0.01$), as shown in Fig.~\ref{fig:energyLandscape_a001}.
Fig.~\ref{fig:small_alpha} shows the orbits in the $(F_1, F_2)$ space for three settings of $r=1, 0.01, 100$ with 20 initial data.
The trajectory bundle for $r=0.01 \ll 1$ contracts and converges to the minimizer $\tt{A110}$, similarly for $r=100 \gg 1$, the bundle contracts and converges to the minimizer $\tt{Z210}$.
The trajectories for $r=1$ initially expand, then gather at a saddle group P.
Subsequently, some trajectories converge directly to $\tt{A110}$, others are attracted to $\tt{A211}$ and then converge to $\tt{A110}$, and there is one trajectory converging directly to $\tt{Z210}$.
To check the tendency of bundle behaviors, we performed similar simulations for 40 different random seeds. Table \ref{tb:statistics_r1} presents the frequency with which initial data trajectories pass through saddle points, detailing their intermediate paths and final destinations.
Here, we define the saddle with the smallest $\ell^2$-error relative to the numerical solution as the nearest saddle. We refer to the saddles as passed through if they satisfy the condition that the Morse index decreases monotonically over time.
From this table, we can see that many orbits (37 initial conditions in this case) pass near the saddle $\tt{C223}$. This overall behavior is consistent with the trajectory bundle observed for the case with 20 random seeds.

There is a sharp contrast between the two cases $r=0.01, 100$ and $r=1$. Uncertainty is reduced when the ratio $r$ deviates significantly from $1$, but the trajectory is very sensitive to the random seed for $r=1$. This can be understood from the view point of slow-fast system. When the ratio $r$ deviates significantly from $1$, one of the variables evolves much faster than the other and reaches quickly a quasi-stationary state, then serves as the environment for the other slow variable as in Fig.~\ref{fig:dynamics}(a)(c).
Therefore, in the latter half of the time evolution, the slow variable $v$ that is close to 0-mode at $t=10.0$ grows into one-hump solution (see Fig.~\ref{fig:dynamics}(a)).
In other words, variables with fast time constants are quickly coarse-grained, i.e., they relax to their one-mode structures, and effectively become the ``environment'' for the remaining variables.
From the perspective of energy decay, when $r \ll 1$, the decay of $F_1$ containing the Ginzburg--Landau energy of $u$ proceeds first, followed by the decay of $F_2$ containing the Ginzburg--Landau energy of $v$ (see Fig.~\ref{fig:flow_r001}).
When $r \gg 1$, the order of decay of $F_1$ and $F_2$ is reversed (see Fig.~\ref{fig:flow_r100}).
Specifically, when $r=0.01$, the orbits move horizontally and pass through $\tt{D008}\to\tt{D206}\to\tt{Z104}\to\tt{Z122p}\to\tt{A110}$, and when $r=100$, the orbits move vertically and pass through $\tt{D008}\to\tt{D026}\to\tt{Z014}\to\tt{Z210}$.
In other words, the bundle of 20 trajectories contracts and behave like a single orbit converging to the same minimizer when $r \ll 1$ or $r \gg 1$.

\begin{figure}[htbp]
  \centering
  \includegraphics[width=1.0\linewidth]{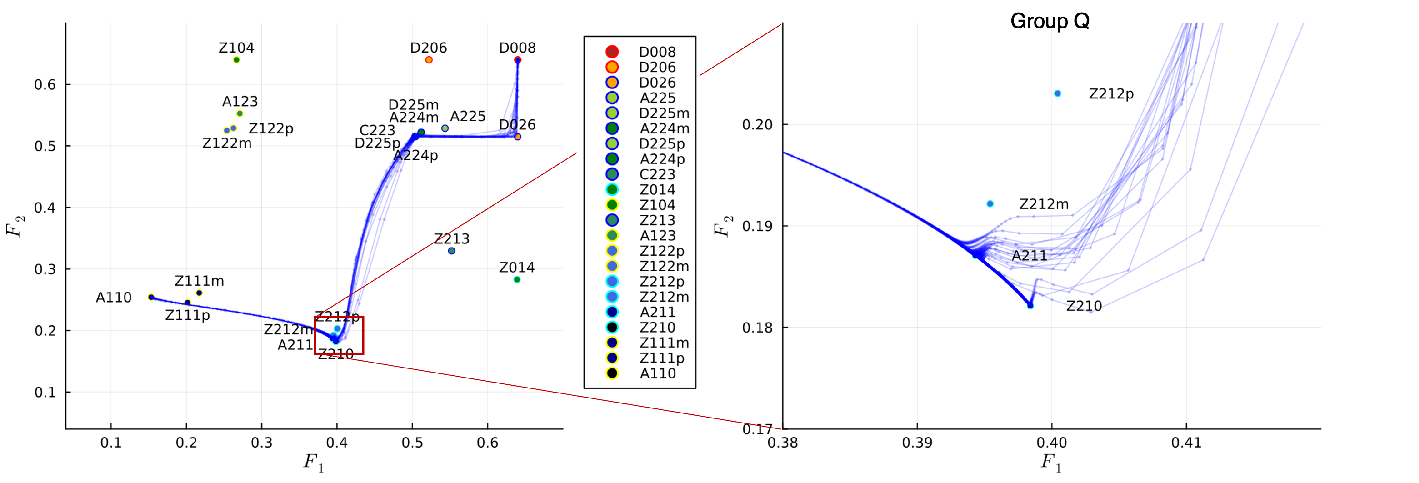}
  \caption{
    Trajectories in the $(F_1,F_2)$ energy space for 20 initial data with $\alpha=0.01$ and $r=1.38$.
    The bundle is thinner than in the case of $r=1$. This is partly because $v$ responds more quickly than $u$, causing most trajectories to evolve nearly vertically at the early stage. They are then drawn toward the saddle sequence $\tt{A224j}\to\tt{C223}\to\tt{A211}$, and finally sorted by the hub saddle $\tt{A211}$ into either $\tt{A110}$ or $\tt{Z210}$. See the main text for details.
  }
  \label{fig:small_alpha_r1.38}
\end{figure}

\begin{table}[htbp]
  \centering
  \resizebox{\textwidth}{!}{
    \begin{tabular}{c|ccccccccccc|c}
      \multirow{17}{*}{Path} & \multicolumn{12}{c}{$\tt{D008}$} \\
      \cline{2-13}
      &  \darrow{7}  &  \darrow{5}  &  \darrow{5}  &  \darrow{3}  &  \darrow{5}  &  \darrow{7}  &  \darrow{7}  &  \darrow{3}  &  \darrow{5}  &  \darrow{1}  &  \darrow{1}  &  \darrow{5}  \\
      &              &              &              &              &              &              &              &              &              & $\tt{A225 }$ & $\tt{D026} $ &              \\
      &              &              &              &              &              &              &              &              &              &  \darrow{5}  &  \darrow{3}  &              \\
      &              &              &              & $\tt{D225m}$ &              &              &              & $\tt{D225p}$ &              &              &              &              \\
      &              &              &              &  \darrow{1}  &              &              &              &  \darrow{5}  &              &              &              &              \\
      &              & $\tt{A224p}$ & $\tt{A224m}$ & $\tt{A224m}$ & $\tt{A224p}$ &              &              &              & $\tt{A224m}$ &              & $\tt{A224m}$ & $\tt{A224p}$ \\
      &              &  \darrow{1}  &  \darrow{1}  &  \darrow{1}  &  \darrow{1}  &              &              &              &  \darrow{5}  &              &  \darrow{1}  &  \darrow{3}  \\
      & $\tt{C223} $ & $\tt{C223} $ & $\tt{C223} $ & $\tt{C223} $ & $\tt{C223} $ & $\tt{C223} $ & $\tt{C223} $ &              &              & $\tt{C223} $ & $\tt{C223} $ &              \\
      &  \darrow{3}  &  \darrow{3}  &  \darrow{3}  &  \darrow{3}  &  \darrow{5}  &  \darrow{7}  &  \darrow{1}  &              &              &  \darrow{3}  &  \darrow{3}  &              \\
      &              &              &              &              &              &              & $\tt{Z212p}$ & $\tt{Z212m}$ &              &              &              & $\tt{Z212m}$ \\
      &              &              &              &              &              &              &  \darrow{3}  &  \darrow{3}  &              &              &              &  \darrow{1}  \\
      & $\tt{A211} $ & $\tt{A211} $ & $\tt{A211} $ & $\tt{A211} $ &              &              &              &              & $\tt{A211} $ & $\tt{A211} $ & $\tt{A211} $ & $\tt{A211} $ \\
      &  \darrow{3}  &  \darrow{3}  &  \darrow{3}  &  \darrow{3}  &              &              &              &              &  \darrow{3}  &  \darrow{3}  &  \darrow{3}  &  \darrow{1}  \\
      &              &              &              &              & $\tt{Z111m}$ &              & $\tt{Z111m}$ & $\tt{Z111p}$ &              &              &              & $\tt{Z210} $ \\
      &              &              &              &              &  \darrow{1}  &              &  \darrow{1}  &  \darrow{1}  &              &              &              &              \\
      & $\tt{A110} $ & $\tt{A110} $ & $\tt{A110} $ & $\tt{A110} $ & $\tt{A110} $ & $\tt{A110} $ & $\tt{A110} $ & $\tt{A110} $ & $\tt{A110} $ & $\tt{A110} $ & $\tt{A110} $ &              \\
      \hline
      Frequency & $18$ & $6$        & $5$          & $2$          & $2$          & $1$          & $1$          & $1$          & $1$          & $1$          & $1$          & $1$          \\
    \end{tabular}
  }
  \caption{
    List of saddles passed through by the orbit at $\alpha=0.01$ and $r=1$, and the frequency of initial data that take that path.
    With one exception, most orbits converge to $\tt{A110}$.
    Furthermore, the fact that many orbits pass through $\tt{C223}$ and $\tt{A211}$ suggests that these two saddle points play a critical role in the dynamics. See the text for details.
  }
  \label{tb:statistics_r1}
\end{table}
In contrast to the above case, when $r \sim 1$, i.e., the value of $r$ is a moderate value that includes the steepest descent, the trajectory bundle behaves in a sensitive way as $r$ varies. For instance, when $r=1.38$, Table \ref{tb:statistics_r1.38} shows that approximately half of the orbits converge to the Janus particle (JP) solution ($\tt{A110}$), and the rest converge to the single particle (SP) solution ($\tt{Z210}$) for $\alpha=0.01$. This makes a sharp contrast with $r=1$ in which
most trajectories converge to $\tt{A110}$ as shown in Table \ref{tb:statistics_r1}. In view of Table \ref{tb:statistics_r1.38}, 37 trajectories out of 40 are classified via two saddles $\tt{C223}$ and $\tt{A211}$. Recall the discussion of Sec. 3.4.2 that $\tt{A211}$ is a hub saddle and $r=1.38$ is close to the critical value in which the stable manifold of $\tt{A211}$ intersects with the unstable manifold of $\tt{C223}$. Therefore it is natural to observe such a splitting around $r=1.38$.
Fig.~\ref{fig:small_alpha_r1.38} shows the detailed behaviors of 20 trajectories in the $(F_1, F_2)$ energy space for the same parameter values. At the initial stage, the trajectory bundle slightly expands but moves down vertically and reaches near $\tt{D026}$ because $v$ evolves faster than $u$ thanks to $r > 1$. Then the bundle is attracted to saddle cluster P and passes through either $\tt{A224p}$ or $\tt{A224m}$. Finally the bundle reaches the saddle cluster Q via $\tt{C223}$ where it is classified as either $\tt{A110}$ or $\tt{Z210}$ by the hub saddle $\tt{A211}$.
As shown in Fig.~\ref{fig:small_alpha_r1.38} and Table \ref{tb:statistics_r1.38}, the trajectories are more effectively controlled when $r$ deviates from 1. First it makes a difference of reaction speeds of $u$ and $v$ at the initial stage. Then the trajectories are attracted to the saddle clusters P and Q, where they are classified by the hub saddles such as $\tt{A211}$.  

\begin{table}[htbp]
  \centering
  \begin{tabular}{c|cccc|ccccccc}
    \multirow{17}{*}{Path} & \multicolumn{11}{c}{$\tt{D008}$} \\
    \cline{2-12}
    &  \darrow{1}  &  \darrow{1}  &  \darrow{1}  &  \darrow{1}  &  \darrow{1}  &  \darrow{1}  &  \darrow{1}  &  \darrow{1}  &  \darrow{1}  &  \darrow{5}  &  \darrow{1}  \\
    & $\tt{D026} $ & $\tt{D026} $ & $\tt{D026} $ & $\tt{D026} $ & $\tt{D026} $ & $\tt{D026} $ & $\tt{D026} $ & $\tt{D026} $ & $\tt{D026} $ &              & $\tt{D026} $ \\
    &  \darrow{5}  &  \darrow{3}  &  \darrow{3}  &  \darrow{3}  &  \darrow{3}  &  \darrow{5}  &  \darrow{1}  &  \darrow{1}  &  \darrow{3}  &              &  \darrow{1}  \\
    &              &              &              &              &              &              & $\tt{D225p}$ & $\tt{D225p}$ &              &              & $\tt{D225m}$ \\
    &              &              &              &              &              &              &  \darrow{1}  &  \darrow{1}  &              &              &  \darrow{1}  \\
    &              & $\tt{A224p}$ & $\tt{A224m}$ & $\tt{A224m}$ & $\tt{A224p}$ &              & $\tt{A224p}$ & $\tt{A224p}$ & $\tt{A224p}$ & $\tt{A224p}$ & $\tt{A224m}$ \\
    &              &  \darrow{1}  &  \darrow{1}  &  \darrow{5}  &  \darrow{1}  &              &  \darrow{7}  &  \darrow{1}  &  \darrow{3}  &  \darrow{1}  &  \darrow{5}  \\
    & $\tt{C223} $ & $\tt{C223} $ & $\tt{C223} $ &              & $\tt{C223} $ & $\tt{C223} $ &              & $\tt{C223} $ &              & $\tt{C223} $ &              \\
    &  \darrow{3}  &  \darrow{3}  &  \darrow{3}  &              &  \darrow{3}  &  \darrow{3}  &              &  \darrow{3}  &              &  \darrow{3}  &              \\
    &              &              &              &              &              &              &              &              & $\tt{Z212m}$ &              &              \\
    &              &              &              &              &              &              &              &              &  \darrow{3}  &              &              \\
    & $\tt{A211} $ & $\tt{A211} $ & $\tt{A211} $ & $\tt{A211} $ & $\tt{A211} $ & $\tt{A211} $ &              & $\tt{A211} $ &              & $\tt{A211} $ & $\tt{A211} $ \\
    &  \darrow{3}  &  \darrow{3}  &  \darrow{3}  &  \darrow{3}  &  \darrow{1}  &  \darrow{1}  &              &  \darrow{1}  &              &  \darrow{1}  &  \darrow{1}  \\
    &              &              &              &              & $\tt{Z210} $ & $\tt{Z210} $ & $\tt{Z210} $ & $\tt{Z210} $ & $\tt{Z210} $ & $\tt{Z210} $ & $\tt{Z210} $ \\
    &              &              &              &              &              &              &              &              &              &              &              \\
    & $\tt{A110} $ & $\tt{A110} $ & $\tt{A110} $ & $\tt{A110} $ &              &              &              &              &              &              &              \\
    \hline
    Frequency & $15$ & $4$        & $3$          & $1$          & $7$          & $5$          & $1$          & $1$          & $1$          & $1$          & $1$          \\
  \end{tabular}
  \caption{
    List of saddles passed through by the orbit at $\alpha=0.01$ and $r=1.38$, and the frequency of initial data that take that path.
    Among the 40 trajectories, 23 converge to $\tt{A110}$ and 17 to $\tt{Z210}$.
    Common pathways include $\tt{D026}\to\tt{C223}\to\tt{A211}$ and $\tt{D026}\to\tt{A224p}\to\tt{C223}\to\tt{A211}$.
    These trajectories, which originate from initial data near $\tt{D008}$ are ultimately sorted by the stable manifold of $\tt{A211}$ (see also Fig.~\ref{fig:C223-A211.eps}).
  }
  \label{tb:statistics_r1.38}
\end{table}

\subsection{Three stages of dynamics}

The purpose of this paper is to clarify ``specifically'' how the sorting mechanism is implemented in the saddle networks. We already see in the previous sections that trajectory bundle is effectively controlled when the $r \ll 1$ or $r \gg 1$. In this section, therefore, we focus on the most sensitive regime that $r$ belongs to the intermediate regime, in which expansion and contraction of trajectory bundle occurs.
In general, there are three stages for the dynamics of the CCH system as time proceeds.
\begin{enumerate}
\item Linear regime: 
  The dispersion relation describes the onset of instability at the constant state $\tt{D008}$.
  As shown in Fig.~\ref{fig:disp_rel_a0_a01}, two modes are dominant in our system size $L=2.56$.
\item Saddle network regime:
  After the onset of linear instability, the trajectory approaches one of the saddle points in the cluster P in most cases. The coarsening process proceeds for either $u$ or $v$ or simultaneously depending on $r$-value.
\item Sorting regime:
  The solution eventually converges to a stationary solution via hub saddle.
\end{enumerate}

In what follows, we consider more detailed mechanism how the expansion and contraction of solution bundle occur along the aforementioned three stages. When $r \sim 1$, one key reason for this deformation is phase misalignment within the linear regime.
In this regime, since the relaxation parameters are nearly the same, both $u$ and $v$ exhibit the same most unstable wave number (two--mode in this case), and during this process, a phase difference generally arises between the two mode solutions of $u$ and $v$, which depends on the random seeds.
Although multiple saddle points exist in the two-hump component (see Fig.~\ref{bifAtlasTeX2.eps}), it is the magnitude of the phase difference that determines which saddle point dominates the next stage of dynamics.
That is, a tiny phase differences at the outset can critically influence the subsequent dynamics.
The dynamics in the second saddle network regime can be classified into the following two types based on the phase difference at the end of the linear regime.
\begin{itemize}
\item {\bf{The case where the phases are aligned}}:\\
  For example, when $u$ and $v$ have the same seed, or when $u$ and $-v$ have the same seed, the orbits often approach $\tt{D225m}$ (in-phase) or $\tt{D225p}$ (anti-phase), respectively.
  However, there are exceptions because general initial data contain various components other than two--mode, even if $u = \pm v$, the orbit may not pass near $\tt{D225m}$ or $\tt{D225p}$, while even if $u \neq \pm v$, it may pass near them. Here we can assume that when $u$ and $v$ are aligned at the end of the linear regime, they are close to $\tt{D225j}$ generically, therefore, the behavior beyond this point depends significantly on the behavior of the unstable manifold of ${\tt D225}j$ (see Table 4 and Table 5). Along the most unstable direction from ${\tt D225}j$, a deformation is induced that maintains left-right symmetry and accelerates the coarse-graining of $v$, leading to the formation of a single peak and ultimately reaching the two-peak solution ${\tt Z212}j\to\tt{Z210}$.

  In the case of the second most unstable eigendirection, $u$ and $v$ are coarsened simultaneously to become a single peak, passing through ${\tt Z111}j$ and finally arriving at the JP solution.
  In the case of the third, fourth, and fifth unstable eigendirection, different situations arise for $\tt{m}$ and $\tt{p}$.
  For the third unstable eigendirection of $\tt{D225p}$ and the fourth unstable eigendirection of $\tt{D225m}$, $u$ coarsens quickly to a single peak, the orbit passes near ${\tt Z122}j$ once, and then $v$ catches up, leading to the JP solution via ${\tt Z111}j$.
  In contrast, for the fourth unstable eigendirection of $\tt{D225p}$ and the fifth unstable eigendirection of $\tt{D225m}$, $v$ coarsens quickly to a single peak, the orbit passes near
  $\tt{Z213}$
  and then reaches the JP solution via ${\tt Z111}j$.
  Regarding the fifth unstable eigendirection of $\tt{D225p}$ and the third unstable eigendirection of $\tt{D225m}$, unlike the other unstable eigenfunctions, which are bilaterally symmetric, it has twofold rotational symmetry.
  Therefore, in this case, the left-right symmetry is immediately broken, the coarsening of $v$ proceeds first, and the orbit passes near $\tt{A211}$.
  Subsequently, in the case of $\tt{m}$, the solution converges to the JP solution $\tt{A110}$, while in the case of $\tt{p}$, the symmetry is restored and the solution converges to $\tt{Z210}$.
  Here, when passing through ${\tt Z111}j$, as in the previous paths, the solution is divided into either the mirror image of polymer AB or BA.
  The sorting mechanism of trajectories by saddles described here depends on the history of the orbits, i.e., it depends on the seed of the initial perturbations, therefore, it is not an easy task to predict
  a priori which instability dominates the dynamics of a given trajectory. However, we can make a list of all possibilities after the linear regime. Possibly we can find a probability distribution of outcomes for an appropriate set of random seeds, but this remains a future work.

  \begin{table}[htbp]
    \centering
    \begin{tabular}{ccr}
      Symmetry & Eigenvalue & Path \\\hline
      $\tt{Z}$ & $1.238$ & $\tt{Z212m}\to\tt{Z210}$ \\
      $\tt{Z}$ & $0.934$ & $\tt{Z111m}\to\tt{A110}$ \\
      $\tt{Z}$ & $0.348$ & $\tt{Z122p}\to\tt{Z111p}\to\tt{A110}$ \\
      $\tt{Z}$ & $0.262$ & $\tt{Z213}\to\tt{Z111m}\to\tt{A110}$ \\
      $\tt{C}$ & $0.014$ & $\tt{A211}\to\tt{Z210}$ 
    \end{tabular}
    \caption{
      Unstable eigendirection of $\tt{D225p}$:
      In particular, the transition from $\tt{A211}$ to $\tt{Z210}$ along the fifth unstable eigendirection represents a symmetry-restoring change and occurs relatively infrequently.
    }
  \end{table}
  
  \begin{table}[htbp]
    \centering
    \begin{tabular}{ccr}
      Symmetry & Eigenvalue & Path \\\hline
      $\tt{Z}$ & $1.243$ & $\tt{Z212p}\to\tt{Z210}$ \\
      $\tt{Z}$ & $1.160$ & $\tt{Z111p}\to\tt{A110}$ \\
      $\tt{C}$ & $0.713$ & $\tt{C223}\to\tt{A211}\to\tt{A110}$ \\
      $\tt{Z}$ & $0.578$ & $\tt{Z122m}\to\tt{Z111m}\to\tt{A110}$ \\
      $\tt{Z}$ & $0.292$ & $\tt{Z213}\to\tt{Z111p}\to\tt{A110}$ 
    \end{tabular}
    \caption{
      Unstable eigendirection of $\tt{D225m}$.
    }
    \label{tb:D225m}
  \end{table}

\item {\bf{The case where the phases are not aligned (generic case) }}(Fig.~\ref{fig:dynamics_energy_flow}):\\
  In most cases,
  the trajectories pass near the saddle point $\tt{C223}$ (see Table 6) and connected to $\tt{A211}$.
  When a perturbation is added to the most unstable direction from $\tt{C223}$, $v$ immediately changes from double peaks to single peak. Here, the term $\beta uv^2$ ($\beta = -0.3$) of the free energy plays a key role: To minimize the energy, most of $v$ shifts toward regions where $u \sim 1$, while $v$ changes sign in regions where $u \sim -1$, resulting in a single-mode profile for $v$. Finally, the orbit proceeds toward the JP solution following the path $\tt{C223}\to\tt{A211}\to\tt{A110}$.
  
  In the second and third unstable eigendirections, the orbit immediately converge to $\tt{A110}$ without passing through $\tt{A211}$.
  Owing to the broken mirror symmetry, the minimizer $\tt{A110}$ can distinguish between two types of polymer positions, AB and BA.
  Although this distinction does not depend on the sign of the unstable eigenfunction, the separation direction is inverted when comparing the first and third unstable eigendirections to the second.
  
  These transition processes change significantly if $r$ increases.
  When
  $r$ slightly
  increases,
  (e.g. $r=1.45$), the sorting at the
  hub saddle $\tt{A211}$ changes to
  $\tt{Z210}$ (the SP solution), as confirmed by numerical experiments (Fig.
  \begin{figure}[htbp]
    \centering
    \includegraphics[width=1.0\linewidth]{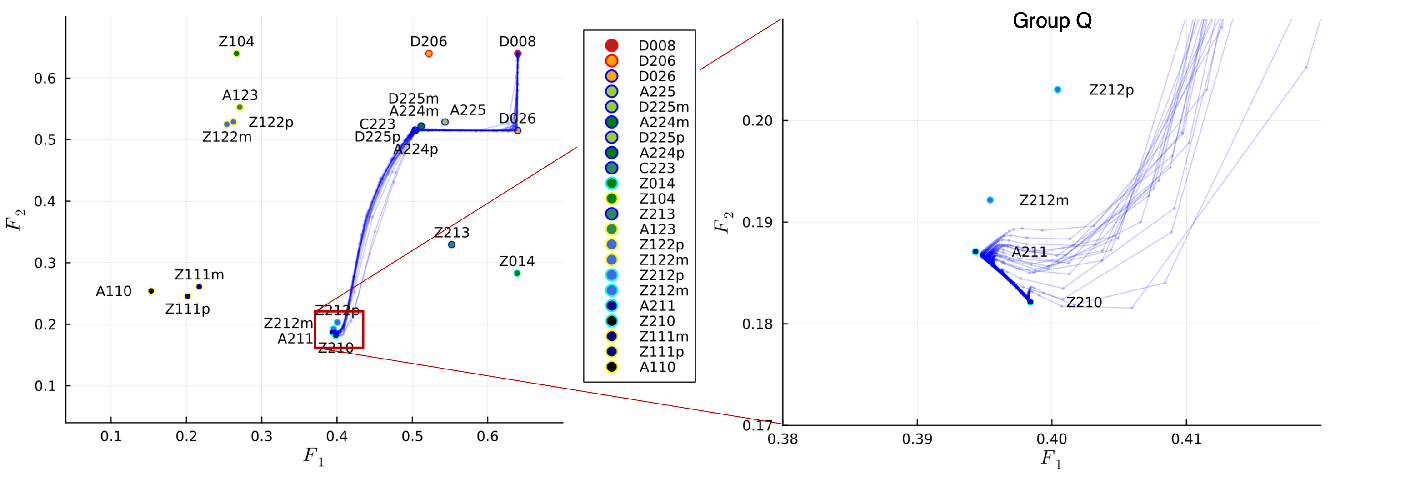}
    \caption{
      Trajectories in $(F_1,F_2)$ energy space for 20 initial data with $\alpha=0.01$ and $r=1.45$.
      Here the ratio $r$ exceeds the critical value $r^*=1.41$ (see Fig.~\ref{fig:C223-A211.eps}). An examination of flow dynamics around Group Q reveals that all orbits converge to $\tt{Z210}$ via the hub saddle $\tt{A211}$.
    }
    \label{fig:small_alpha_r1.45}
  \end{figure}
  \ref{fig:small_alpha_r1.45}).
  Schematic diagram of this change is shown
  shown in Fig.~\ref{fig:C223-A211.eps}.
  $\tt{C223}$ and its neighbor are on the basin of the minimizer $\tt{A110}$ when $r \ll r^*$, and hence, orbits with initial data near $\tt{C223}$ converge to $\tt{A110}$ when $r=1$.
  On the other hand, for $r \gg r^*$, orbits with initial data near $\tt{C223}$ are attracted to $\tt{Z210}$ since $\tt{C223}$ and its neighbor are on the basin of $\tt{Z210}$.
  When $r \approx r^*$, both of the above cases occur.
  \begin{table}[htbp]
    \centering
    \begin{tabular}{ccr}
      Symmetry & Eigenvalue & Path \\\hline
      $\tt{A}$ & $2.205$ & $\tt{A211}\to\tt{A110}$ \\
      $\tt{A}$ & $1.235$ & $\tt{A110}$ \\
      $\tt{A}$ & $0.314$ & $\tt{A110}$ 
    \end{tabular}
    \caption{
      Unstable eigendirection of $\tt{C223}$.
    }
  \end{table}
\end{itemize}
In summary, in generic cases where the phases are not aligned, almost all solutions converge to the JP solution $\tt{A110}$ via $\tt{C223}$ when $r=1$.
However, this also depends on the ratio $r$, and when $r$ is slightly increased (e.g., $r=1.45$), sorting often reverses at the saddle point $\tt{A211}$ and is attracted to the SP solution $\tt{Z210}$.
Incidentally, even if the phases are not aligned, it is possible that the solutions approach a saddle point with a slightly misaligned phase, such as ${\tt A224}j$.
In this case, as well, it is numerically verified that the solutions pass through $\tt{C223}$ and frequently converge to $\tt{A110}$.
In this sense, the saddle point $\tt{A211}$ is a ``hub saddle'' that determines whether the orbit converges to the JP solution $\tt{A110}$ or the SP solution $\tt{Z210}$, therefore, the heteroclinic connection $\tt{C223}\to\tt{A211}$ is important.
Moreover, the ratio $r$ critically determines which side of the stable manifold of the hub saddle $\tt{A211}$ the trajectory evolves toward the JP solution or SP solution.

\section{Conclusion and outlook}
\subsection{Why relaxation parameters matter?}
The free energy defined by (\ref{eqn:energy}) has huge number of local minimizers and saddles in 2D and 3D spaces (see, for instance, \cite{Nishiura_Landscape2D}, \cite{Avalos2016} and \cite{Avalos+2024} and reference therein). Therefore, it is not a priori clear that what kind of local minimizer is selected even for a fixed initial data and energy. On the other hand, in experiments, a class of exotic morphologies is typically observed with high probability, once all the experimental settings such as pressure and initial concentration are fixed. This discrepancy seems to come from the lack of knowledge how the above experimental conditions are incorporated in the theoretical computation, in fact, those conditions are not linked to the parameters of the righthand side of the free energy. The remaining possibility is the ratio of relaxation parameters $r = \tau_u / \tau_v$ that is usually unknown a priori, therefore, we typically adopt the steepest descent method to trace the trajectory from a given initial condition, namely the ratio is equal to one. 
This seems to be a natural selection, because the energy decreases most efficiently when the trajectory is orthogonal to the contour line. However, this is not always the case in real experiments, for example, precipitation is accelerated as pressure is decreased so that the formation of particles becomes faster than before, which suggests that the relaxation parameter in front of $u$ variable should be smaller than that of $v$. Namely, the ratio can be changed according to the experimental conditions. This approach, grounded in the aforementioned observation, utilizes the adjustment of relaxation parameter ratios to concurrently explore minimizers and saddle points, in a manner consistent with experimental setups.

\subsection{Expansion and contraction of trajectories}
The bundle of trajectories with various random seeds contract or expand depending on the ratio $r$ of relaxation parameters as was discussed in Sec. 4. A remarkable thing is that the bundle becomes almost a single trajectory independent of random seeds when $r\ll 1$ and $r\gg 1$, namely it itinerates unique set of saddles and minimizers independent of fluctuation, which is consistent with experimental setups adopted for Platon solids. This contraction was confirmed numerically at least in one-dimensional space and all the minimizers can be found through the exploration of the extreme cases of $r$, which means that only two simulations are enough to detect the targeted minimizers. This is useful to avoid a brute force approach like carpet bombing with various initial data with random seeds and to give us an edge of global landscape before searching more precisely. It is expected in higher dimensional space that similar behaviors can be observed, in fact, very exotic like polyhedral morphologies in 3D have been found in such a special regime of $r$.

\subsection{Infinite dimensional slow manifold}
When there is a large disparity in the time scales of $u$ and $v$, it is natural to reformulate the system as a slow--fast system. 
Using the ratio $r$, the model system (\ref{eqn:CCH}) is equivalent to the following slow system:
\begin{equation}
  \left\{
  \begin{aligned}
    & r \frac{\partial u(t,x)}{\partial t} =  \frac{\partial^2}{\partial x^2}\Big[ -\epsilon_u^2 \frac{\partial^2}{\partial x^2} u - u + u^3 + \alpha v + \beta v^2 \Big]
    \ \text{for}\ t>0, x\in[0,L),
      \\
      & \frac{\partial v(t,x)}{\partial t} =  \frac{\partial^2}{\partial x^2}\Big[ -\epsilon_v^2 \frac{\partial^2}{\partial x^2} v - v + v^3 + \alpha u + 2\beta uv \Big]
      \ \text{for}\ t>0, x\in[0,L).
  \end{aligned}
  \right.
  \label{eqn:CCH_r}
\end{equation}
\noindent
On the other hand, Introducing the fast time $\eta = t/r$, \eqref{eqn:CCH_r} becomes

\begin{equation}
  \left\{
  \begin{aligned}
    & \frac{\partial u(t,x)}{\partial \eta} = \frac{\partial^2}{\partial x^2}\Big[ -\epsilon_u^2 \frac{\partial^2}{\partial x^2} u - u + u^3 + \alpha v + \beta v^2 \Big]
    \ \text{for}\ t>0, x\in[0,L),
      \\
      & \frac{\partial v(t,x)}{\partial \eta} = r \frac{\partial^2}{\partial x^2}\Big[ -\epsilon_v^2 \frac{\partial^2}{\partial x^2} v - v + v^3 + \alpha u + 2\beta uv \Big]
      \ \text{for}\ t>0, x\in[0,L).
  \end{aligned}
  \right.
  \label{eqn:CCH_fast}
\end{equation}
\noindent
Taking the formal limit $r \rightarrow 0$, $v$ becomes a function independent of $\eta$ and $u$ evolves according to the above single equation that is the fast equation. The whole dynamics starting from $(u,v) = (0,0)$ consists of the fast part \eqref{eqn:CCH_fast} with $v=0$ followed by the slow one \eqref{eqn:CCH_slow} defined below. The fast motion denoted by $u^{*}(t,x;v \equiv 0)$ is exactly the same as the single Cahn-Hilliard equation for $u$, therefore it converges to a one-hump solution $U^{*}$. After this process, the slow variable $v$ starts to evolve according to the formal limit $r \rightarrow 0$ of \eqref{eqn:CCH_slow}, namely

\begin{equation}
  \left\{
  \begin{aligned}
    & 0 = \frac{\partial^2}{\partial x^2}\Big[ -\epsilon_u^2 \frac{\partial^2}{\partial x^2} u - u + u^3 + \alpha v + \beta v^2 \Big]
    \ \text{for}\ t>0, x\in[0,L),
      \\
      & \frac{\partial v(t,x)}{\partial t} = \frac{\partial^2}{\partial x^2}\Big[ -\epsilon_v^2 \frac{\partial^2}{\partial x^2} v - v + v^3 + \alpha u + 2\beta uv \Big]
      \ \text{for}\ t>0, x\in[0,L).
  \end{aligned}
  \right.
  \label{eqn:CCH_0}
\end{equation}
\noindent
The first equation is a fourth-order stationary problem and $(U^{*}, 0)$ is a solution. Suppose that it can be solved locally with respect to $u$ as $u=H^{-1}(v)$, then the second equation becomes

\begin{equation}
  \frac{\partial v(t,x)}{\partial t}
  = \frac{\partial^2}{\partial x^2}\Big[ -\epsilon_v^2 \frac{\partial^2}{\partial x^2} v - v + v^3 + \alpha H^{-1}(v) + 2\beta H^{-1}(v)v \Big]
  \ \text{for}\ t>0, x\in[0,L).
    \label{eqn:CCH_slow}
\end{equation}
\noindent
This describes the slow evolution of $v$ on the infinite dimensional slow manifold defined by $u=H^{-1}(v)$ starting from $(U^{*}, 0)$.
Solving this equation, and assume the orbit converges to the steady state $V^{*}=V^{*}(U^{*})$ and denote the trajectory by the same notation $v^{*}=v^{*}(\eta,x;U^{*})$. The union of the trajectory $u^{*}(t,x;v \equiv 0)$ and $v^{*}=v^{*}(\eta,x;U^{*})$ is the zeroth approximation of the fast motion and slow motion. In fact, Fig.~\ref{fig:flow_r001} suggests that this is the case. First, the orbit proceeds horizontally ($v$ remains almost constant there) and $u$ reaches a quasi steady state close to $(U^{*}, 0)$, then slow variable $v$ starts to evolve and reaches the local minimizer $\tt{A110}$ eventually. The same thing happens when $r$ is extremely large as shown in Fig.~\ref{fig:flow_r100}.
In general, however, the construction of slow manifold is not an easy task, in fact, it is not clear how the above inverse $u=H^{-1}(v)$ can be defined globally and what is the dynamics on it. These problems remain as a future challenge.

\subsection{Splitting method}
When the relaxation parameters of $u$ and $v$ differ significantly, it is natural to apply the operator-splitting method (see \cite{Blanes_splitting_2024} and reference therein) and reformulate the system as a slow--fast computational scheme for the coupled Cahn--Hilliard equations. Although the exact form of the slow manifold is difficult to determine as indicated in the item 5.3, this parameter contrast enables a simplified numerical strategy ($r \ll 1$ case):
\begin{enumerate}
\item Fast-step: Solve the fast equation for several time steps until $u$ reaches a quasi-steady state.

\item Slow-step: Keeping $u$ fixed, integrate the slow equation for several steps to advance $v$.

\item Iterate these two steps until convergence to a steady state.

\end{enumerate}
This approach significantly accelerates computation compared to solving the fully coupled system in tandem. In the full system, the allowable time step is constrained by the fast variable, enforcing an unnecessarily small step for the slow variable due to the small ratio $r$. In contrast, the split scheme permits each variable to be computed with its own appropriately sized time step, thereby reducing total computational time. For example, if the calculation with $r=100$ is performed on the full system, the splitting method approximates the result in about $1/100$ of the calculation time.

\begin{figure}[htbp]
  \centering
  \begin{tabular}{c}
    \begin{subfigure}[t]{\linewidth}
      \centering
      \includegraphics[width=\linewidth]{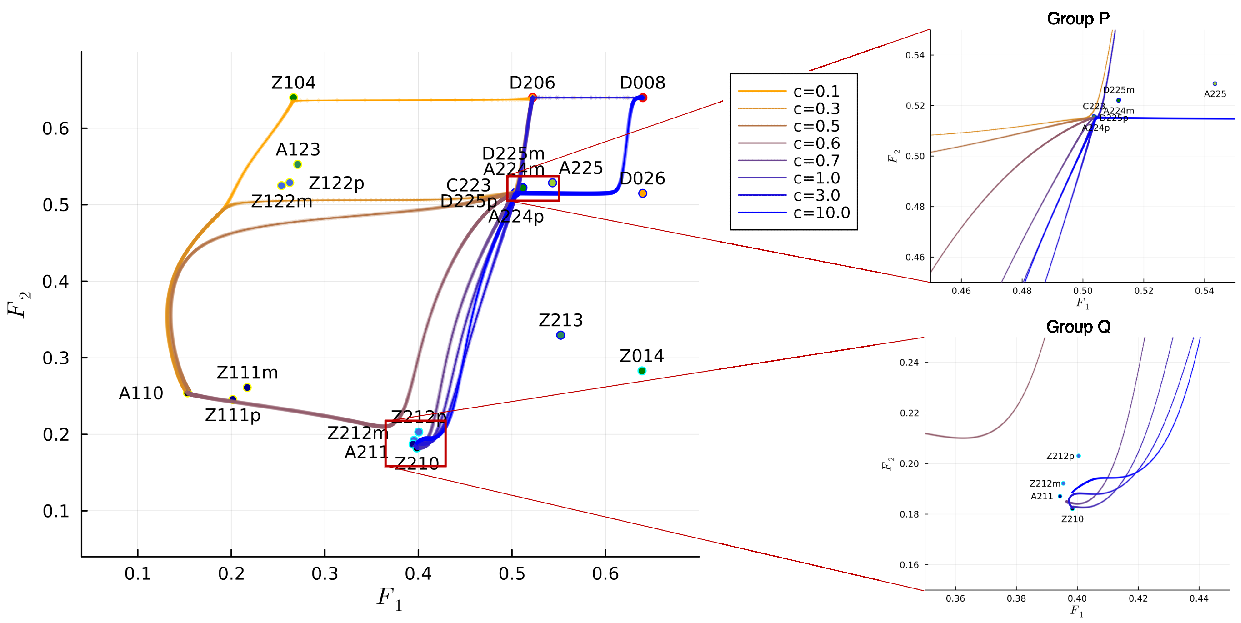}
      \subcaption{$0.1 \leq c \leq 10$}
      \label{fig:wide_c}
    \end{subfigure}
    \\[6ex]
    \begin{subfigure}[t]{\linewidth}
      \centering
      \includegraphics[width=\linewidth]{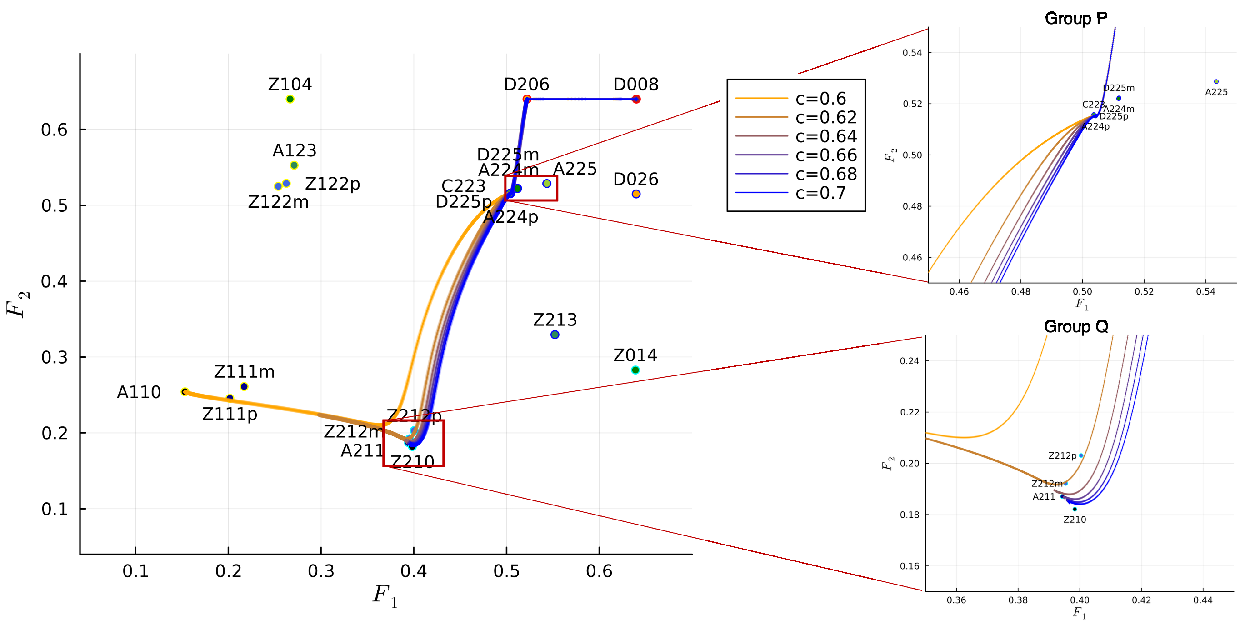}
      \subcaption{$0.6 \leq c \leq 0.7$}
      \label{fig:tipping_c}
    \end{subfigure}
  \end{tabular}
  \caption{
    Trajectories for various values of $c$ with a fixed initial datum in the $(F_1,F_2)$ energy space.
    The rate $r(t)$ is given by $r=ct+0.01$ and grows linearly with respect to time.
    When $c$ is small, the orbit converges to $\tt{A110}$, while for larger values of $c$, it converges to $\tt{Z210}$, with a switching point around $c\sim 0.66$.
    The pathway $\tt{D206}\to$ Group P $\to$ Group Q, shown in panel (b), does not appear when the ratio $r$ is fixed and time-independent. 
  }
  \label{fig:time_dependent_ratio}
\end{figure}

\subsection{Time-dependent ratio $r(t)$}

Another interesting approach is to consider the case in which the ratio $r$ is a function of time.
As already observed, when $r \ll 1$ or $r \gg 1$, the bundle of trajectories contract and converge to one of the local minimizers independent of random noise. 
However, in a more general setting, there appear many other local minimizers that are located outside of the extreme region $r \ll 1$ or $r \gg 1$. 
Again, it may be possible to find a good $r$-value to hit a new morphology by trial and error, but this may be not so efficient. 
Another idea is to consider the trajectory with $r(ct)$ being a function of time. 
Here $c$ is the rate of change. Changing the rate in an appropriate way and collecting itinerancy of the associated trajectories, it might be possible to sweep the phase space efficiently and find a exotic morphologies with high probability. 
In fact, Fig.~\ref{fig:wide_c} shows the case in which $r(ct)$ changes in a linear manner $r=ct + 0.01$ and how the trajectory behaves as $c$ varies from $0.1$ to $10.0$ for a fixed initial datum. When $c$ is small, the trajectories explore the upper-left region of $(F_1,F_2)$ space and gradually descend as $c$ increases. All of them eventually converge to $\tt{A110}$. As $c$ increases further, 
Fig.~\ref{fig:tipping_c} illustrates the switching behavior of trajectories within the range $0.6 \leq c \leq 0.7$. 
When $c$ is below $0.64$, the trajectory approach the minimizer $\tt{A110}$. However, as $c$ nears $0.66$, a tipping point is reached, and for $c$ greater than $0.68$, the trajectories instead settle into the minimizer $\tt{Z210}$. 
In other words, tipping is not always bad, rather quite useful to explore the landscape globally.
Such tipping-like transitions frequently occur in more general settings. We believe that employing a time-dependent ratio $r(t)$ provides a powerful tool for exploring the global energy landscape and identifying desired morphologies.

\subsection{Higher dimensional space}
As was mentioned in Sec. 1, the present study was strongly motivated by the exotic morphologies of BCPs as in Fig.~\ref{fig:yabu} (see also \cite{Avalos+2024}). Especially it was a surprising and counterintuitive result that Platonic polyhedra were found in such a small scale both in experiments and simulations. The smallness of ratio $r$ means that particle formation is a fast process, which is consistent with the specific experimental conditions such as lower pressure or high concentration of initial polymers that accelerate the precipitation of nanoparticles. In such a regime, various polyhedra have been found with high probability and the number of faces is increased as the size of the particle becomes larger. It is a big challenge to explore the atlas of saddles and find the network among them in 2D and 3D, even though it is quite complicated as a whole.

\section*{Acknowledgment}
Y.N. thanks Hiroshi Yabu for his valuable comments and suggestions. This work was partially supported by JSPS KAKENHI Grant Number 23K17653, JP23K19003.

\bibliographystyle{unsrt}
\bibliography{PhysicaD_2025_delAbst_fixed_edit}

\end{document}